\documentclass[10pt]{article}
\usepackage{psfig}
\usepackage{graphicx}
\usepackage{psfrag}   
\usepackage{latexsym}    
\usepackage{amsmath} 
\usepackage{amssymb}

\newtheorem{lemma}{Lemma}

\newtheorem{theor}{\large\bf Theorem}
\newtheorem{coroll}{\large\bf Corollary}

\def\FF{\hbox to 8.33887pt{\rm I\hskip-1.8pt F}}
\def\NN{\hbox to 9.3111pt{\rm I\hskip-1.8pt N}}
\def\PP{\hbox to 8.61664pt{\rm I\hskip-1.8pt P}}
\def\QQ{\rlap {\raise 0.4ex \hbox{$\scriptscriptstyle |$}}
{\hskip -4.5pt Q}}
\def\RR{\hbox to 9.1722pt{\rm I\hskip-1.8pt R}}
\def\ZZ{\hbox to 8.2222pt{\rm Z\hskip-4pt \rm Z}}
\def\ZZZ{Z\!\!\!Z}

\newcommand{\resetsect}{\setcounter{section}{1}}

\newcommand{\resetequ}{\setcounter{equation}{0}}
\newcommand{\be}{\begin{equation}}
\newcommand{\ee}{\end{equation}}
\newcommand{\bqa}{\begin{eqnarray}}
\newcommand{\eqa}{\end{eqnarray}}
\newcommand{\ba}{\begin{array}}
\newcommand{\ea}{\end{array}}

\newcommand{\no}{\nonumber}

\newcommand{\lp}{\left (}
\newcommand{\rp}{\right )}
\newcommand{\qed}{\hfill \rule {1ex}{1ex}}
\newcommand{\al}{\alpha}

\newcommand{\bt}{\beta}

\newcommand{\de}{\delta}
\newcommand{\vep}{\varepsilon}

\newcommand{\la}{\lambda}

\newcommand{\si}{\sigma}

\newcommand{\La}{\Lambda}

\newcommand{\De}{\Delta}

\begin{document}

\centerline{\Large \bf Density of states for  Random Band Matrix}
\vskip 0.2cm

\centerline{ \bf M. Disertori$^a$\footnote{supported by NSF grant
 DMS 9729992}, H. Pinson$^b$, T. Spencer$^a$}
\vskip 0.2cm

\centerline{a) Institute for Advanced Study}
\centerline{Einstein Drive, Princeton}
\centerline{NJ 08540,  USA}

\centerline{b) University of Arizona}
\centerline{ Tucson, Arizona 85721, USA }
 \vskip 1cm

\begin{abstract}
By applying the supersymmetric approach we rigorously prove  smoothness
of the averaged density of states for a three dimensional random
band matrix ensemble, in the limit of infinite volume and fixed band width.
We also prove that the resulting expression for the density of states 
coincides with the Wigner semicircle with a precision $1/W^2$, for $W$
large but finite.
\end{abstract}

\section{Introduction}
\resetequ

Random Matrix Theory (RMT) has proved to be relevant in
the study of several physical models. It was initially applied to 
the study of resonance spectra of complex nuclei and later to the study of the
quantum properties of weakly disordered
conductors, and the  spectral properties of quantum systems 
which are chaotic in their classical limit \cite{M}\cite{AOS}. 
RMT also appears  in other fields, such as statistics, number theory
and random permutations. See for example \cite{BA}\cite{KS}\cite{J}
for recent developments.

In this article we study the density of states for a 
class of Hermitian random matrices $H_{ij}$ whose elements 
are Gaussian with mean zero and covariance 
\be
\left \langle H_{ij} H_{kl} \right \rangle =
\de_{jk}\  \de_{il} \ J_{ij} \ .
\label{covar}\end{equation}
In the classical case of GUE, Gaussian unitary ensembles, the
indices $i$ and $j$ range from 1 to $N$ and 
$J_{ij}=1/N$. For this case the density of states  (DOS) is given by 
Wigner's famous  semicircle law 
\begin{equation}
\rho_{SC}(E) =
\begin{cases}
\frac{1}{\pi}  \sqrt{1-\frac{E^2}{4}} & |E|\leq 2 \\
  \quad 0 & |E|> 2
\end{cases}
\label{sc}\end{equation}
in the limit $N\uparrow\infty$. Our analysis will focus on band random matrices
for which the indices $i$, $j$ range over a box $\La\cap\ZZZ^d$
and $J_{ij}$ is small when $|i-j|$ is larger then some fixed band width $W$.
As we let $\La\uparrow\ZZZ^d$ the spectral properties of such matrices should 
be quite similar to that of a random Schr\"odinger operator
on a lattice $\ZZZ^d$ given by 
\be
H = -\De + \la V(j)
\label{schr}\end{equation}
where $V(j)$ are independent random variables and $\De$ is the discrete Laplacian.
For example in one dimension the spectra of the random Schr\"odinger and the 
random band matrix are pure point with exponentially decaying eigenfunctions
\cite{PF} \cite{AS}.
This is called localization. In two or more dimensions, localization 
also holds for energies outside some interval
depending on $d$, $\la$, $W$.
Thus band random matrices are a way of interpolating between classical 
random matrix ensembles (GUE or GOE) and  random Schr\"odinger. 

The goal of this article is to obtain detailed information about the density 
of states for a special class of random band matrices in 3 dimensions.
We shall consider energies at which extended rather than localized states are 
expected.
More precisely let $i$, $j\in \La\cap \ZZZ^3$, $\La$ a set of cubes
of side $W$, and define
\begin{equation}
J_{ij} := \
\lp\frac{1}{- W^2 \De + 1 }\rp_{ij} 
\simeq\  \tfrac{1}{4\pi W^2} \tfrac{1}{(1+|i-j|)} e^{-\frac{|i-j|}{W}}
\label{Jdef}\end{equation}
where $\De$ is the Laplacian with periodic boundary conditions  
in the volume $\La$ and $W$ is large but fixed.
Our estimates are valid uniformly in the size of  $ \La\subseteq\ZZZ^3$.
The average density of states is given by 
\be
\left \langle \rho_\La(E)\right \rangle =  -\frac{1}{\pi}  
\lim_{\vep\downarrow 0}{\rm Im}\
\left \langle  \lp \frac{1}{E+i\vep -H}\rp_{00}\right \rangle \ .
\end{equation}
Note that as $\La\uparrow\ZZZ^3$, the density of states $\rho(E)$ does not
depend on the configuration with probability one. 
The derivative of $\rho_\La(E)$ is 
\be
\frac{d}{dE} \left \langle \rho_\La(E)\right \rangle  = 
 \frac{1}{\pi}  \lim_{\vep\downarrow 0}{\rm Im}\
\sum_{x\in \La} R(E+i\vep; 0,x) 
\end{equation}
where 
\be
R(E+i\vep; 0,x)  =  \left \langle
\lp \frac{1}{E+i\vep -H}\rp_{0x} \lp \frac{1}{E+i\vep -H}\rp_{x0}        
\right \rangle
\end{equation}
Note  that for $x\neq 0$ 
\be 
\left \langle
\lp \frac{1}{E+i\vep -H}\rp_{0x}        
\right \rangle = 0 
\end{equation}
because of the symmetry $H_{ij}\rightarrow - H_{ij}$.

Our main result is that for large $W$ and $E$ inside the interval 
$[-2,2]$, $\left \langle \rho_\La(E)\right \rangle$ equals the Wigner
semicircle distribution  (\ref{sc})  plus corrections of order $W^{-2}$.
Moreover $R(x)$ decays exponentially fast and $\rho(E)$ is
smooth in $E$. These results hold for fixed $W$ and are uniform in $\vep$ as 
$\vep\downarrow 0$ and in the volume as $\La\uparrow \ZZZ^3$. 
See Theorem 2.1 for a precise statement. When $|E|>2+O(W^{-1})$ we expect that
$\rho(E)$ is smaller than any power of $W^{-1}$ and that localization holds.

For random Schr\"odinger operators given by (\ref{schr}) we have the classic bound by 
Wegner, $\rho(E) \leq const\, \la^{-1}$ for small $\la$. This estimate
is far from optimal since for small $\la$ we expect the density of states 
to approach that of the Laplacian. Unfortunately there are no uniform bounds 
on $\rho(E)$  as $\la\rightarrow 0$ or estimates on the smoothness of $\rho(E)$
unless either the  distribution of $V$ is Cauchy 
(in which case the density of states can be 
explicitly computed) or $E$ lies in an interval for  which localized states 
are proved to exist. 

Note that for both the random Schr\"odinger and the random band 
matrix ensembles it is conjectured that for $d=3$
\be
\pi \sum_x \left \langle
\left | \lp  \frac{1}{E+i\vep -H}\rp_{0x}\right |^2  
 \right \rangle \ e^{ixp} \simeq \frac{\rho(E)}{Dp^2 +\vep}
\label{powerl}\end{equation}
where $D$ is the diffusion constant. Here $E$ must be inside $[0,4d]$
for the case of random Schr\"odinger or inside  $[-2,2]$ for our band matrix ensemble,
 and both $W^{-1}$ and $\la$ are small.
This paper does not address this important conjecture. Instead we are using the 
phase oscillations of the Green's functions to obtain exponential decay 
for $R(x)$.

To establish our results on Green's functions we use the supersymmetric 
formalism of K. Efetov \cite{Ef} \cite{Ef1} which has its roots 
in earlier work by Wegner
  \cite{W} \cite{SW}. We recommend the survey article of Mirlin  \cite{M}
and also the paper of Mirlin and Fyodorov \cite{MF} which studies random
band matrices in 1 dimension. In the mathematics literature,
A. Klein studied the density of states using supersymmetric methods  \cite{Kl}
but only at energies where localization holds.

The supersymmetric method enables one  to explicitly average the Green's function over 
the randomness. 
This technique involves the use of both real and anticommuting variables. However
when we perform our estimates all anticommuting variables are integrated out so that
the resulting integrals is just over real variables.
As a result of this averaging, the problem is converted into a 
problem in statistical mechanics whose action has approximately the form
\be
{\cal A}(\phi) = \sum_{j\in\La} W^2 \lp\nabla\phi \rp^2(j) + U(\phi)(j)
\end{equation}
where the potential $U$ is a function of the field $\phi(j)$ and has two saddle points. 
In some
respects this problem looks like a double well $\phi^4$ interaction. 
A more careful analysis 
of the integral over $\phi$ shows that one saddle dominates and it yields the
Wigner semicircle distribution.
The second saddle is suppressed by a determinant as we shall explain later. 
The large parameter $W$ ensures that the integral is governed by the saddle 
and its Gaussian
fluctuations. 
There are similar integrals which appear for random Schr\"odinger operators,  
however the  path integral is much more oscillatory and we can not yet control 
them unless there are long range correlations in the $V(j)$. 

The average of $\left | (E+i\vep -H)^{-1}_{0x} \right |^2$ can also 
be calculated with the 
supersymmetric formalism but the statistical mechanics is now more complicated.
Instead of two saddle points there is a non compact saddle manifold 
and fluctuations have massless modes which are responsible for the power law 
 (\ref{powerl}). 

The remainder of this article is organized as follows. In Sec.~2 we give
a precise statement of our results. In Sec.~3 we use the supersymmetric formalism 
to convert averages of the Green's function to a model in statistical mechanics.
The advantage of this representation is that for large $W$ we see that the
integral is dominated by two saddle points. These saddle 
points and their Hessians are discussed in Sec.~4. The following section is 
devoted to obtaining our results  in a box $\La\subseteq \ZZZ^3$ of side $W$.
In the last section we  show that the analysis in the box can be extended 
to $\ZZZ^3$ using a variant  of the cluster expansion.

\paragraph{Notation} As in the paper we will need to insert 
many constants in the different bounds we will denote by $K$
any large positive constant, independent from $W$ and $\La$, and
by $c$ any small positive constant  independent from $W$ and $\La$.
These constants need  not  be the same in different estimates.
Also we will  sometimes  
use the symbol $\lesssim$ to indicate that there is a constant
factor $K$ on the right side of the inequality ($\lesssim$ stands for $\leq K$)
without writing $K$ explicitly.

\paragraph{Acknowledgments.} We thank Rowan Killip for many discussions
and suggestions related to this paper. These discussions lead us
to improve the proof,  in particular by 
introducing Brascamp-Lieb inequalities.

\section{Model}
\resetequ

As we said in the introduction, we consider the set ${\cal H}$  
of Hermitian matrices $H$ with entries  $i,j\in \La\subset \ZZZ^{\ d}$, $d>0$.
From (\ref{covar}) we see that the  probability density is
\be
P(H) = \prod_{\substack{ij\in \La \\ i<j}} 
\frac{dH^{}_{ij} dH^*_{ij} }{2\pi J_{ij}}
e^{-\frac{|H_{ij}|^2}{J_{ij}}}
 \prod_{i\in \La} \frac{dH_{ii} }{\sqrt{2\pi J_{ii}}}
e^{-\frac{H_{ii}^2}{2 J_{ii}}}
\label{proba}\end{equation}
where $<$ is an order relation on $\La$ and $J$ is defined
in (\ref{Jdef}).
With these definitions $ {\cal H}$ is a set of hermitian random band 
matrices with band width $W$. Note that in $d=1$ for $\La=[1,N]$
we have $|\La|=N$ and  $J_{ij}= N^{-1} \exp[-|i-j|/N ]$
which is very close to  GUE.
For any function of $H$ $F(H)$
we define the average $\langle F(H)\rangle$ as
\be
\langle F(H)\rangle = \int dH\, P(H)\, F(H)
\end{equation}
We study the  averaged density of states   
${\bar\rho}_\La(E)$:
\be
{\bar \rho}_\La(E) =: -\frac{1}{\pi} \lim_{\vep\rightarrow 0^+}
{\rm Im}\left \langle\frac{{\rm Tr} }{|\La|} \frac{1}{E+i\vep -H}
\right \rangle
= -\frac{1}{\pi}  \lim_{\vep\rightarrow 0^+}{\rm Im}\
\left \langle G^{+}_{00}\right \rangle
\end{equation}
where ``Im'' indicates the imaginary part and
$G^{+}$ is the retarded  Green's function:
\be
G^{+} :=  \frac{1}{E+i\vep -H} \ .
\end{equation}

In the following we 
restrict to $d=3$ and we consider energies well
inside the spectrum. For technical reasons we also avoid the energy $E=0$.
Therefore we consider only energies in  the interval
\be
{\cal I}=: \{ E \, : \, |E|\leq 1.8 \ {\rm and}\ |E|>\eta\}
\label{interv}\end{equation}
with $\eta>0$. 
We assume that our region $\La$ is a union of cubes in $\ZZZ^3$ of side $W$.
The paper is devoted to the proof of the  following theorem

\begin{theor}
For $d=3$, there exists a value $W_0$ such that for all $W\geq W_0$
the averaged density of states  ${\bar \rho}_\La(E)$ 
is smooth in $E$, in  the interval  ${\cal I}$, uniformly in
$W$ and $\La$ (hence also in the limit $\La\uparrow\ZZZ^3$): 
\be
\left | \partial_E^n \, {\bar \rho}_\La(E)\right | \ \lesssim\  C_n\  \ \forall\ 
n< n_0(W)
\label{smooth}\end{equation}  
where $\lim_{W\uparrow\infty} n_0(W)= \infty$. 
Moreover  ${\bar \rho}_\La(E)$ is the semicircle law with a precision 
$1/W^2$.
\be
\left |  {\bar \rho}_\La(E)
-\rho_{SC}(E)
 \right |
\lesssim \ \frac{1}{W^2}
\end{equation}
where $\rho_{SC}(E)$ is the semicircle law $\rho_{SC}$ 
defined in (\ref{sc}).
Note that the first equation (\ref{smooth}) means in particular
that for $x\neq 0$
\begin{equation}
R(x)= \left |  \left\langle
G^+_{0x}\, G^+_{x0} \right\rangle_{\vep=0}\right |\  
\lesssim \ \frac{1}{W^3}\, e^{-c\, \frac{|x|}{W}}
\label{expd}\end{equation}
\end{theor}

\paragraph{Outline of the paper}

In Sec.~5 we  establish Theorem~1 on  a cube $\La$ of side $W$.
We use the supersymmetric formalism 
to write $\langle   G^+_{00}\rangle$ as a
functional integral where a saddle point analysis
can be performed. 
Actually there are two saddles. For $d=3$  one  saddle is suppressed
by a factor $e^{-W}$ (note that this is not true for  $d\leq 2$). 
The fluctuations around the saddle are  
controlled using small probability arguments while 
the integral near the dominant saddle is estimated by 
a Brascamp-Lieb inequality \cite{BL}\cite{Sp}.

Sect.~6  is devoted to the cluster expansion which enables us
to analyze the limit  $\La\uparrow \ZZZ^3$. The cluster expansion
expresses  $\langle G^+_{00}\rangle$ as a sum over finite volume contributions 
$Y\subset \ZZZ^3$ which are again unions of cubes of side $W$. 
We show that large $Y$ terms give small contributions to
 $\langle G^+_{00}\rangle$. This expansion also enables us to prove the bound
on $R(x)$.

\section{Supersymmetric approach}
 \resetequ

In this section we shall use the algebraic formalism of  supersymmetry to
express our average Green's function in terms of a functional integral,
which, apart form a determinant, is local. Let $J$ be given by (\ref{Jdef}).

\vskip 0.2cm

\begin{lemma}
The averaged Green's function can be written
as
\begin{align}
\left\langle G^+_{00}\right\rangle = &
 \int da\,  db\   
\exp\left[-\frac{1}{2} (a^T J^{-1} a
+  b^T J^{-1} b ) \right] 
 \left [ \prod_{i\in \La} \frac{(E_\vep-ib_i)}{(E_\vep-a_i)}
\right ] \no\\
& \times  \frac{1}{(E_\vep-a_0)}  \det [ J^{-1}-F(a,b)-F'(a_0,b_0)]  
\label{formula} \end{align}
where $a$ and $b$ are  vectors whose components  $a_i$, $b_i$ 
($i=1,...,|\La|$) 
are real variables and  $a^T$ and $b^T$ are the corresponding
transposed vectors.  We defined $E_\vep=E+i\vep$ and 
$F(a,b)$ and $F'(a_0,b_0)$  are  matrices with elements 
\be
F(a,b)_{ij} := \de_{ij}  \frac{1}{(E_\vep-a_i)(E_\vep-ib_i)}
\label{Fdef}\end{equation}
\be
F'(a_0,b_0)_{ij} := \de_{i0}  \de_{j0}   \frac{1}{(E_\vep-a_0)(E_\vep-ib_0)}
\label{Fdef1}\end{equation}
Note that each $a_i$ has a pole at $a_i=E+i\vep$ while 
$b_i$ has no singularity (as it appears only in the numerator).
This expression is then well defined only for $\vep >0$.

By the same technique we obtain a similar formulas  for 
$\langle G^+_{0j}\, G^+_{j0}\rangle$ and in general for
$\langle G^+_{0j_1}\,G^+_{j_1j_2}\,...\, G^+_{j_n0}\rangle$.
\end{lemma}

\paragraph{Remarks} Note that if we omit the observable, that
is we omit  $(E_\vep-a_0)^{-1}$
and $F'$ in (\ref{formula}), we are actually computing $\left\langle 1\right\rangle=1$,
thus
\begin{equation}
1 = \int da\,  db\   
\exp\left[-\frac{1}{2} (a^T J^{-1} a
+  b^T J^{-1} b ) \right] 
 \left [ \prod_{i\in \La} \frac{(E_\vep-ib_i)}{(E_\vep-a_i)}
\right ]  \det [ J^{-1}-F(a,b)]  \ .
\label{one} \end{equation}

\paragraph{Proof}
Note that the Green's function can be written as a functional integral:
\be
G^+_{kl} = -i\int dS^* dS \ \exp\left [i S^+ (E_\vep -H) S
\right ]\,
\det[-i (E_\vep -H)]\, S_k S^*_l \
\end{equation}
where the determinant is the normalization factor and we defined 
\be
S = \lp\ba{c}
S_1 \\
\vdots \\
S_{|\La|} \\
\ea \rp \qquad S^+ = \lp S_1^*, \hdots, S_{|\La|}^* \rp
\end{equation}
and $S_1$,...$S_{|\La|}$ are complex bosonic fields.
In order to  insert all the $H$ dependence in the argument 
of the exponential we introduce integrals over fermionic fields:
\be
\det[-i (E_\vep -H)]  = \int d\chi^* d\chi  
\ \exp\left [ i \chi^+ (E_\vep -H) \chi\right ]
\end{equation}
where
\be
\chi = \lp\ba{c}
\chi_1 \\
\vdots \\
\chi_{|\La|} \\
\ea \rp \qquad \chi^+ = \lp \chi_1^*, \hdots, \chi_{|\La|}^* \rp
\end{equation}
and $\chi_1$,...$\chi_{|\La|}$ are fermionic complex fields.
Therefore we can write
\be
G^+_{kl} = -i\int d\Phi^* d\Phi\ \exp\left[i \Phi^+ (E_\vep -H) 
\Phi\right]  \,S_k S^*_l  
\end{equation}
where we have introduced the superfields $\Phi_1$,...
$\Phi_{|\La|}$ ($i=1,\cdots, |\La|$) 
\be
\Phi_i = \lp\ba{c}
S_i \\
\chi_i \\
\ea \rp \qquad \Phi_i^+ = \lp S_i^*, \chi_i^* \rp
\end{equation}
 These superfields can be seen
as components of a supervector $\Phi$
\be
\Phi = \lp\ba{c}
\Phi_1 \\
\vdots \\
\Phi_{|\La|} \\
\ea \rp \qquad \Phi^+ = \lp \Phi_1^+, \hdots, \Phi_{|\La|}^+ \rp
\end{equation}
We adopted the conventions in the review by Mirlin \cite{M}
We summarize supersymmetric formalism and notation  in App.~A.

Now we can perform the average over $H$:
\be
\left\langle \exp\left [-i \Phi^+ H \Phi\right ]\right \rangle =  
\exp\left [-\frac{1}{2}
\sum_{ij} J_{ij} (\Phi_i^+\Phi^{}_j) 
(\Phi_j^+\Phi^{}_i) \right ]
\end{equation}
To convert this quartic interaction into a quadratic one we
perform a Hubbard-Stratonovich transformation:
\be
\sum_{ij} J_{ij} (\Phi_i^+\Phi^{}_j) 
(\Phi_j^+\Phi^{}_i) = \sum_{ij} [ A_i J_{ij} A_j - 
 B_i J_{ij} B_j - 2 P^*_i   J_{ij} P_j ]
\end{equation}
where
\be
A_i = S^*_i S_i, \qquad B_i = \chi^*_i \chi_i, \qquad
P_i = S^*_i \chi_i, \qquad P^*_i = S_i \chi^*_i
\end{equation}
and there is no sum over $i$. Note that $A_i$ and $B_i$ are commuting 
variables while $P_i$ and $P^*_i$ are anticommuting ones. Now
\bqa
 \exp\left[ -\frac{1}{2} A^T J A \right]  &=& (2\pi)^{-\frac{|\La|}{2}}
\int 
\frac{\prod_{i\in \La} da_i}{\sqrt{\det J}}\ 
 e^{  -\frac{1}{2} a^T J^{-1} a - i a^T A } \\
\exp\left[   +\frac{1}{2} B^T J B\right]     &=&  (2\pi)^{-\frac{|\La|}{2}}
 \int 
\frac{\prod_{i\in \La} db_i}{\sqrt{\det J}}\ 
  e^{-\frac{1}{2} b^T J^{-1} b + b^T B }  \no\\
\exp\left[    - P^+ J P \right]   &=& (2\pi)^{|\La|} \int 
\frac{\prod_{i\in \La} d\rho^*_i d\rho_i}{\det J^{-1}}\ 
  e^{   - \rho^+ J^{-1} \rho - i \rho^+ P -i P^+ \rho }\no 
\eqa
where $a_i$, $b_i$ are real bosonic fields and $\rho_i$ is a  
complex fermionic field for any $i=1,...,|\La|$. 
Therefore
\be
\left\langle \exp\left [-i \Phi^+ H \Phi\right ]\right \rangle 
\ =\  \int da\,  db\,  d\rho^*\,  d\rho \ e^{ -\frac{1}{2} (a^T J^{-1} a
+  b^T J^{-1} b + 2  \rho^+ J^{-1} \rho ) -i \Phi^+ R \Phi} 
\end{equation}
where 
\be
\Phi^+ R \Phi := \sum_{i=1}^{|\La|} \Phi_i^+ R_{i} \Phi_i , \qquad 
R_i := \lp \ba{cc}
a_i & \rho^*_i \\
\rho_i & i b_i \\ 
\ea \rp
\end{equation}
$R_i$ is actually a supermatrix, containing both bosonic and 
fermionic variables. For such a matrix we can define 
the notion of transpose, complex conjugate, determinant
and trace and it can be shown that the usual properties of the vector and 
matrix algebra hold.  We summarize the notations
in App.~A.
Using this formalism we have
\be
\left\langle\int d\Phi^* d\Phi\ e^{ i \Phi^+ (E_\vep -H) 
\Phi}\, S_k^{} S^*_l \right\rangle \,= \, \frac{\de_{kl}}{-i} \ \lp 
\frac{1}{E_\vep-R_k}\rp_{11} 
\prod_{i} \frac{1}{{\rm Sdet} (E_\vep-R_i)}  
\end{equation}
where 
\be
{\rm Sdet} (E_\vep-R_i) = \frac{(E_\vep-a_i)}{(E_\vep-ib_i)} 
\left [ 1 - \rho^*_i\rho_i^{}  \frac{1}{(E_\vep-a_i)(E_\vep-ib_i)}  \right ]
\end{equation}
\be
\lp \frac{1}{E_\vep-R_k}\rp_{11} =  \frac{1}{(E_\vep-a_k)}
\left [ 1 - \rho^*_k\rho_k^{}  \frac{1}{(E_\vep-a_k)(E_\vep-ib_k)} 
 \right ]^{-1}
\end{equation}
Therefore
\bqa
&&\hspace{-0.5cm} \left\langle G^+_{00}\right\rangle =   
\int da\,  db\,   d\rho^*  d\rho\
e^{-\frac{1}{2} (a^T J^{-1} a
+  b^T J^{-1} b + 2  \rho^+ J^{-1} \rho )}\,
\left [ \prod_{i\in \La} \frac{(E_\vep-ib_i)}{(E_\vep-a_i)}
\right ] 
\no\\
&&\times  \, \frac{1}{(E_\vep-a_0)} 
\left [ 1 - \tfrac{ \rho^*_0\rho_0^{}}{(E_\vep-a_0)(E_\vep-ib_0)} 
 \right ]^{-1}
\prod_{i\in \La}
\left [ 1 - \tfrac{\rho^*_i\rho_i^{} }{(E_\vep-a_i)(E_\vep-ib_i)}  
\right ]^{-1}
\eqa

The integration over the fermionic fields can
be performed  exactly. 
Using the property: $\rho_i^2=(\rho^*_i)^2=0$  $\forall i$
we observe that
\be
\left [
1 - \tfrac{ \rho^*_i\rho_i^{}}{(E_\vep-a_i)(E_\vep-ib_i)}\right ]^{-1}
= 
 \exp\left [  \tfrac{\rho^*_i\rho_i^{}}{(E_\vep-a_i)(E_\vep-ib_i)}
\right ]
\end{equation}
therefore the integration over $\rho$ and $\rho^*$  reduces to the following
expression
\be
  \int d\rho^* d\rho \  e^{- \rho^+ [J^{-1}+F(a,b)+F'(a_0,b_0)] \rho }
\ =\  \det\, [ J^{-1}-F(a,b)-F'(a_0,b_0)]   
\end{equation}
where $F(a,b)$  and $F'(a_0,b_0)$  are defined in (\ref{Fdef}-\ref{Fdef1}).
We obtain then the expression (\ref{formula}).
\vskip 0.2cm

\section{Saddle point analysis}
\resetequ

In this section we shall deform the  integral (\ref{formula})
over $a_j$  and $b_j$ so that they pass through certain complex 
saddle points. 
If we ignore the determinant in  (\ref{formula2}) and the
kinetic term, we show that the  resulting integrand has a double well structure,
with the two wells of the same height. 
In the  Sec.~5.1.2 (Lemma~6) we will see that the determinant
actually suppresses one of the two saddles by a factor $e^{-W}$.
\vskip 0.2cm

\paragraph{Saddle points}

Observing the integrand in (\ref{formula}) we remark that  
the factor $-W^2\De$ in $a^T J^{-1}a+b^T J^{-1}b$
forces the fields $a$ and $b$  to be approximately constant. Therefore
if we ignore the determinant, the leading 
contribution to the integrand (hence also to the saddle structure) is then 
\be
e^{-\frac{|\La|}{2} (a^2 +  b^2 )} 
 \left [\frac{(E_\vep-ib)}{(E_\vep-a)}
\right ]^{|\La|}  =  \left [e^{-[f_1(a)+f_2(b)]}\right ]^{|\La|}
\label{domc}\end{equation}
where  the fields $a$ and $b$ are
constant ($a_i=a$,  $b_i=b$  for all $i$) and we defined 
\begin{align}
f_1(a) & = \frac{a^2}{2} + \ln (E-a)\no\\
f_2(b) & =  \frac{b^2}{2}  - \ln (E-ib).
\label{domc1}\end{align} 
Note that in this approximation the saddle points for 
GUE and Random Band Matrix are the same.
The critical points of $f_1$ and $f_2$ are given by
\bqa
a_s  &=& \quad  {\cal E}_r \pm i  {\cal E}_i  
\label{saddlep}\\
b_s &=& -i {\cal E}_r \pm {\cal E}_i  
\no
\eqa
where we defined
\be
{\cal E} := {\cal E}_r - i {\cal E}_i:=  \frac{E}{2} - i 
\sqrt{1-\frac{E^2}{4}}  \ .
\label{energy}\end{equation}
Note that  ${\cal E}$ satisfies 
\be
E- {\cal E}  = {\cal E}^*  , \quad  {\cal E}{\cal E}^*  =1  
\quad \forall\, |E|<2  \ .
\label{prop1}\end{equation}

\paragraph{Spectrum}

Note that, if $|E|<2$ the saddle $a_s$, $b_s$ have non zero
imaginary parts even as  $\vep\downarrow 0$. 
For $|E|\geq 2+ O(W^{-1})$ we expect that the density of states is 
smaller than any power of $W^{-1}$, for $W$ large.

\paragraph{Contour deformation}

We  deform the integration contour in order to
pass through a saddle point.  To avoid crossing 
the  pole $a_i=E_\vep$ we have to pass through the saddle
$a_s= {\cal E}$.  On the other hand
the choice for   $b_s$ is arbitrary, as there is no pole
in $b$, but it turns out (see Sec.~5.1) that  $b_s=-i{\cal E}$ is
the dominant contribution. 
Note that
\be
f''_1({\cal E}) = f''_2(-i{\cal E}) = (1- {\cal E}^2) \ .
\end{equation}
Hence the Hessian at this saddle point is 
\be
B^{-1} = -W^2 \De + (1- {\cal E}^2)
\label{B2}
\end{equation}

\begin{lemma} 
We perform inside  (\ref{formula}) the translation 
\bqa
a_j &\rightarrow & a_j \,+\, {\cal E} 
\label{shift}\\
b_j &\rightarrow & b_j \,-i {\cal E} \quad \forall\, j\in \La\no
\eqa
and take  the limit 
$\vep\downarrow 0$. The integral can then  be written as
\be 
\left\langle G^+_{00}\right\rangle_{\vep =0} =  
 \int d\mu_B(a,b)\,  
  \det [ 1 + (D+D'_0) B]\,   
e^{V'_0+\sum_{j\in \La} V_j }
\label{formula2} \end{equation}
where the measure  $d\mu_B(a,b)$ (\ref{B1}-\ref{B2})  has covariance $B$
given by (\ref{B2}). 
The factor $\exp [\sum_j V_j]$ (\ref{vertex1}) 
is what remains in the exponential after the Hessian
has been extracted, $\det[1+DB]$  (\ref{vertex2}) corresponds to 
$\det[J^{-1}+F]$ after
the normalization factor $\det B^{-1}$ has been extracted. Finally
$\exp V'_0$ and $D'_0$  (\ref{vertex3}-\ref{vertex4})   
are the contributions from the observable.

More precisely we define the measure as 
 $d\mu_B(a,b)=  d\mu_B(a)\,  d\mu_B(b)$ with  
\be
 d\mu_B(a) = \sqrt{\det B}\ 
e^{-\frac{1}{2} (a^T B^{-1} a)}\, , \qquad 
d\mu_B(b) = \sqrt{\det B}\ 
e^{-\frac{1}{2} (b^T B^{-1} b)}\ , 
\label{B1}\end{equation}
and $B$ is the Hessian around the saddle, defined in (\ref{B2}) .
The normalization factor for the measure has been extracted 
from the determinant.
The interactions are given by $V_j = V_j(a_j) + V_j(b_j)$ 
and $D_{ij} = \de_{ij} D_i$,  
where 
\bqa
V_j(a_j) &=&
 \int_0^1 dt\, (1-t)^2 \frac{a_j^3}{( {\cal E}^* -t a_j)^3} \no\\
V_j(b_j) &=&  -\int_0^1 dt\, (1-t)^2 
\frac{\lp ib_j\rp^3}{({\cal E}^*-t ib_j)^3} 
\label{vertex1}\eqa
\bqa
 D_i &=&  [ {\cal E}^2 -  F(a+{\cal E},b-i{\cal E})_{ii} ]
= {\cal E}^2 - \frac{1}{({\cal E}^* - a_i) ({\cal E}^* -ib_i )} 
\label{vertex2}\\
  &=& - \int_0^1 dt \,\left [   
\frac{a_i}{( {\cal E}^* -t a_i)^2 ( {\cal E}^* -ib_i t)} + 
\frac{ib_i}{({\cal E}^* -t a_i) ({\cal E}^* -ib_i t)^2}
\right ] \ .\no
\eqa
Finally the contributions from the observable are given by
$V'_0$ and $D'_0$ where
\be
V'_0 = - \ln ({\cal E}^*-a_0)  \label{vertex3}
\end{equation}
\begin{equation}
(D'_0)_{ij} = - F'(a_0+{\cal E},b_0-i{\cal E})_{ij}
= -\de_{i0} \de_{j0}\  
\frac{1}{({\cal E}^* - a_0) ({\cal E}^* -ib_0 )} \ .
\label{vertex4}
\end{equation}
\end{lemma}

The proof is a straightforward change of variables and a reorganization of
the resulting expression.
Note that for any $|E|<2$ there is no pole in $a$ as the factor $E-{\cal E}-a_i$
is always at a distance at least ${\cal E}_i$ 
from zero. 

For the special value $E=0$,  
a singularity in $b_i=1$ seems to appear from the
factor $1/i(1-b_i)$ in the argument of the determinant. 
This is not a real singularity as
there is the same factor  in the numerator
outside the determinant. Nevertheless to avoid additional
technical problems we  avoid  
$E=0$ in the following. This is the reason why we chose 
$\eta>0$ in ${\cal I}$ in Theorem 1.

\paragraph{Properties of the Hessian.}

The Hessian  $B^{-1}$ (\ref{B2}), which is the covariance of the  Gaussian measure 
after the translation, has now a complex  mass term:
\be
 (1 - {\cal E}^2) =  2 \lp 1-\frac{E^2}{4}\rp +
i E \sqrt{ 1-\frac{E^2}{4}} =: m_{r}^2 + i m_{i}^2 \ .
\end{equation}
Note that for $|E|<2$ the real part $m_r^2$ is positive and this
ensures the convergence of the integral.
In the following, as  we will need to treat in a different way
the real and imaginary part of $B^{-1}$,  we  
introduce the real covariance $C$ 
\be
C := \frac{1}{- W^2 \De +  m_r^2 }
\label{cdef}\end{equation}
therefore 
\be
B^{-1} = C^{-1} + i  m_i^2 \ .
\end{equation}
Note that $C$ is positive as a quadratic form  and pointwise.
In momentum space $C$ is written as
\begin{align}
\hat{C}(\vec{k}) =& \frac{1}{W^2} 
\frac{1}{2\sum_{i=1}^d \lp 1-\cos k_i\rp + \lp m_r/W \rp^2}
\\ 
 & k_i= 2\pi \frac{n_i}{|\La|^{1/d}}, \quad 
n_i= 0,\hdots, |\La|^{1/d}-1 \ .
\no
\end{align}
When $|\La|\uparrow\infty$, $k_i$ becomes a continuum 
variable $k_i\in [0,2\pi]$.
The spatial decay depends on the dimension. In the particular case of $d=3$
\begin{equation}
0< C_{ij} \lesssim 
\frac{1}{W^2(1+|i-j|)} \ e^{-|i-j| \frac{m_{r}}{W}} \ .
\label{Cdecay}\end{equation}
The covariance $B$ has the same expression as $C$, but
with an imaginary term in the mass. It is easy to prove
that $B$ decays in the same way as $C$.
\begin{equation}
\left |B_{ij}\right| \lesssim 
 \frac{1}{W^2(1+|i-j|)} \ e^{-|i-j| \frac{m_{r}}{W}} \ .
\label{Bdecay}\end{equation}

\paragraph{Properties of the interaction}

\begin{figure}
\centerline{\psfig{figure=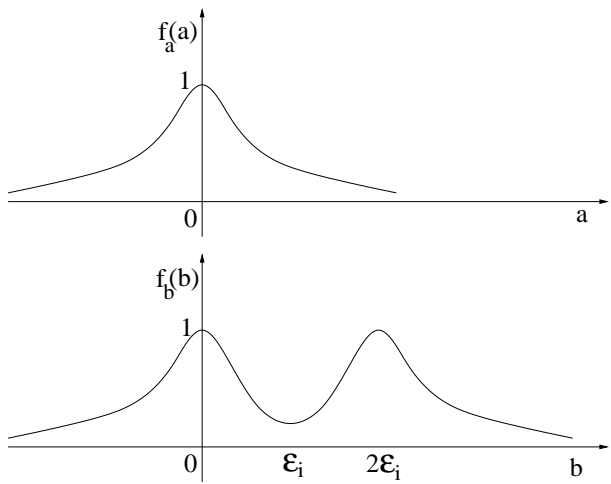,width=8cm}}
\caption{behavior of  $F_1(a)$ and $F_2(b)$  }
\label{inter}
\end{figure}

After the translation the functions $f_1$, $f_2$ introduced in
(\ref{domc1}) become
\begin{align}
f_1(a) &= \  -\frac{1}{2}\lp 1 
- {\cal E}^2\rp a_j^2+  V_j(a_j)   \no\\
f_2(b) & =:\  -\frac{1}{2}\lp 1 
- {\cal E}^2\rp b_j^2+  V_j(b_j)  
\end{align}
Note that, after the translation there also  constant factors
arising from $f_1$ and $f_2$ which cancel.
In the following we will insert absolute values in the integral,
in order to obtain our estimates. We then have to study the behavior of
\be
F_1(a) =: \left |e^{-f_1(a)}\right | \qquad 
F_2(b) =: \left |e^{-f_2(a)}\right |
\end{equation}
It is easy to prove that for $|E|\leq 1.8$   $F_1(a)$
has  only one maximum, in $a=0$, of height 1 (see Fig.~\ref{inter}).
Note that when  $ 1.8< |E|\leq 2$ 
zero is no longer the maximum of $F_1(a)$ and this is why we restrict $E$ to 
${\cal I}$ given by (\ref{interv}). Nevertheless, there is still a single
saddle point  so we expect that by suitable deformation of the contour we should
be able to extend our result to the interval $|E|\leq 2 - O(W^{-1})$.

On the other hand, for any value of $|E|< 2$,  $F_2(b)$  has two maxima, 
which do correspond to the two saddles, one in
$b=0$ and one in $b=2{\cal E}_i$. Both 
maxima have height 1 (Fig.~\ref{inter}). 
We will see in the next section that the second maximum 
 is suppressed by a factor $e^{-W}$
from the determinant appearing in (\ref{formula2}).

\section{Finite volume estimate}
\resetequ

We prove now Theorem~1 in a fixed cube  $\La$ of side $W$, 
with $0\in \La$. We prove  the boundness of ${\bar\rho}_\La(E)$ 
in Theorem~2,  
then in Theorem~3 we prove the  bounds on the derivatives
and on $|{\bar\rho}_\La(E)-\rho_{SC}|$: these bounds  follow by the same technique
used for the bounds on   ${\bar\rho}_\La(E)$, with some slight modifications.

\begin{theor} 
For $\La$ as above,    there exists a value $W_0$ such that for all $W\geq W_0$
and for all $E\in {\cal I}$, where ${\cal I}$ is defined in (\ref{interv}),
the averaged density of states  ${\bar \rho}_\La(E)$ 
is  bounded uniformly in $W$ and $\La$
\be
\left | {\bar\rho}_\La(E)  \right | \leq \  K
\label{boundrho1}\end{equation}
\end{theor}

\begin{theor} 
For $\La$ as above,    there exists a value $W_0$ such that for all $W\geq W_0$
and for all $E\in {\cal I}$, where ${\cal I}$ is defined in (\ref{interv}),
we have
\begin{align}
\left |\partial_E^n {\bar\rho}_\La(E)  \right |& \leq   \ 
C_n \qquad \forall\, n<n_0(W)\\
\left | {\bar\rho}_\La(E)-\rho_{SC}(E)  \right |&  \leq \  \frac{K}{W^2}
\end{align}
uniformly in $\La$ and $W$.
\end{theor}

\subsection{Proof of Theorem 2}

Inserting the absolute values in the expression
(\ref{formula2}) we have
\be 
\left |\left\langle G^+_{00}\right\rangle_{\vep =0} \right | 
\leq \int \left |d\mu_B(a,b)\right |\, \left | \det[1+(D+D'_0)B] \right |\, 
  \left |e^{V'_0+\sum_{j\in \La} V_j } \right |
\end{equation}
The absolute values of $d\mu_B$ and $\det[1+(D+D'_0)B]$ are bounded 
through Lemma~3 and 4 respectively. 

\begin{lemma} 
The total variation of the complex measure is bounded by
\begin{equation}
 |d\mu_{B}(a,b)|\  \lesssim \ e^{O\frac{|\La|}{W^3}} \  d\mu_{C}(a,b)
\label{measure}\end{equation} 
\label{mbound}
\end{lemma}

\paragraph{Proof}
The measure   $d\mu_{B}(a,b)$ can be written as
\begin{equation}
 |d\mu_{B }(a,b)|\ =\ \left |
\frac{\det  B^{-1}}{\det  C^{-1}}
\right |\  d\mu_{C}(a,b)
\end{equation} 
where the determinants are the normalization
factors for the two measures and can be written as
\begin{equation}
 \left |
\frac{\det  B^{-1}}{\det C^{-1}}
\right |\ =\   \left |
\det  1+i m_i^2 C 
\right |\  
\end{equation}
Note that, for any normal  matrix $A$, with  ${\rm Tr} \,A^+A< \infty$,
the following inequality is true 
\begin{equation}
  \left |
\det (1 + A) \right |\ 
\leq \  \left | e^{{\rm Tr}\, A }
 \right | \   e^{\frac{1}{2}\, {\rm Tr}\, A^+A}
\label{detb}\end{equation}
In our case $A= i\de C$, therefore ${\rm Tr}\, A$ is imaginary, and 
the norm of the first exponential is one. The second exponent
gives 
\begin{equation}
 {\rm Tr}\, A^+A = m_i^4 {\rm Tr}\, C^2 =
  m_i^4  \sum_{i,j\in \La} \frac{1}{W^4 |i-j|^2} e^{-\frac{m_r|i-j|}{W}}
\leq  \frac{m_i^4 }{m_r^2} \sum_{i\in \La }  \frac{1}{W^3} 
\end{equation}
The bound in (\ref{measure}) then follows.
\qed
\vskip 0.2cm

\begin{lemma} 
The determinant of $1+(D+D'_0)B$ is bounded by
\begin{equation}
 \left |\det [1+(D+D'_0)B]\right |\, \leq \, 
 K^{\frac{|\La|}{W^3}} \,
\left | e^{{\rm Tr}\, (D+D'_0) B} \right |
\end{equation} 
\end{lemma}

\paragraph{Proof} The proof is obtained by applying 
(\ref{detb}) and repeating the same 
arguments as in the Lemma above. Note that we applied 
$\sup_{ab} |D(a,b)+D'_0(a,b)|=K$ 
for some constant $K$ independent from $W$.
\qed
\vskip 0.3cm

\noindent Applying the Lemmas above we have 
\be 
\left |\left\langle G^+_{00}\right\rangle_{\vep =0} \right | 
\leq  K^{\frac{|\La|}{W^3}} 
\int d\mu_C(a,b)\, \left | e^{{\rm Tr}\, (D+D'_0) B} \right |\, 
  \left |
e^{\sum_{j\in \La} V_j } \right |
\label{formula3} \end{equation}
where we bounded $|\exp(V')|=|{\cal E}^*-a_0|^{-1}\leq K$,
and $V'$ in defined in (\ref{vertex3}) .

\paragraph{Partitioning the domain of integration}

In order to distinguish small field and large
field regions we partition the integration 
domain by inserting 
\be 
1=\sum_{k=1}^5 \chi[I^k]
\label{partition} \end{equation}
 as follows
\begin{align}
&\int d\mu_C(a,b)\, \left | e^{{\rm Tr}\, (D+D'_0) B} \right |\, 
  \left |
e^{\sum_{j\in \La} V_j } \right |\  =\  
 \sum_{k=1}^5 T_k \no\\
& T_k\ =: \ \int d\mu_C(a,b)\, \left | e^{{\rm Tr}\, (D+D'_0) B} \right |\, 
  \left |
e^{\sum_{j\in \La} V_j } \right |\ \chi[I^k] 
\label{part}
\end{align}
where $\chi[I^k]$ is the characteristic function of the set $I^k$
and 

\begin{align}
I^1 & =   \left \{a,b \ :\,  |a_j|,|b_j-b_{j'}|\leq 
 \frac{1}{W^{\frac{1}{8}}}\ \forall j,j'\in \La\, {\rm and}\,
|b_0|\leq \frac{2}{W^{\frac{1}{8}}}
\right \}\no\\
I^2 & =   \left \{a,b \  :\, |a_j|,|b_j-b_{j'}|\leq 
 \frac{1}{W^{\frac{1}{8}}}\ \forall j,j'\in \La\, {\rm and}\,
|b_0-2{\cal E}_i|\leq \frac{2}{W^{\frac{1}{8}}}
\right \} 
\end{align} 
\begin{align}
I^3 & =   \left \{a,b\ :\,  b_j\in \RR \,\forall j\, {\rm and}\, 
\exists\ j\in \La  \ {\rm s.t.}\ |a_j|>
 \frac{1}{W^{\frac{1}{8}}} \right \}
\\
I^4 &=   \left \{a,b \ :\,  |a_j|\leq 
 \frac{1}{W^{\frac{1}{8}}}\ \forall j\in \La\, {\rm and}\, 
\exists\ j,j'\in \La\ {\rm s.t.}\ |b_j-b_{j'}|> \frac{1}{W^{\frac{1}{8}}}
\right \} \no\\
I^5 &=   \left \{a,b \ :\, |a_j|,|b_j-b_{j'}|\leq 
 \frac{1}{W^{\frac{1}{8}}}\ \forall j,j'\in \La\, {\rm and}\,
|b_0|, |b_0-2{\cal E}_i| > \frac{2}{W^{\frac{1}{8}}}
\right \}\no
\end{align}

\paragraph{Small field region}
The first two  intervals correspond to the  {\em small field region}.
$T_1$ is the leading contribution and corresponds to
the case when all  $a$ fields and all $b$ fields  are near zero.
In this case  the interacting terms of the measure do not
destroy the log convexity of the Gaussian $d\mu_C$, 
therefore we can apply a 
Brascamp-Lieb inequality \cite{BL}\cite{Sp} which states

\paragraph{Brascamp-Lieb Inequality:}{\em Let
\be
d\mu_H(x)  =: dx_1...dx_N \frac{1}{Z(H)} e^{-\frac{1}{2}H(x)}
\end{equation}
where $x=(x_1,...,x_N)\in \RR^N$,  $H(x)$ is a positive function
symmetric under $x\rightarrow -x$, and the partition function is
\be
Z(H) =: \int  dx_1...dx_N  e^{-\frac{1}{2}H(x)}
\end{equation}
Then if $H''\geq C^{-1} > 0$ the following inequalities hold
\begin{align}
& \int d\mu_H(x) \   |x_i|^n \leq 
\int d\mu_C(x)\   |x_i|^n  \qquad n>0 \label{bl1}\\
& \int d\mu_H(x)  \   e^{ ( f, x)} \leq 
\int d\mu_C(x)\   e^{ ( f, x)} 
\label{bl2}\end{align}
where  $d\mu_C(x)$
is the free measure with covariance $C$,  $f$ is any vector in $\RR^N$,  
and $(f,x)=\sum_i f_i x_i$.
}\vskip 0.2cm

The second term  corresponds to the case when
all the  $a$ fields are near zero and 
all  the $b$ fields are  near the second saddle $2{\cal E}_i$
(see Fig.~\ref{inter}). In this case we bound the interaction
(trace and $V_j$ factors) by  sup norm. The large contributions
are now suppressed by a small $\exp[-W]$ factor, from the trace bound.

\paragraph{Large field region}
The last three intervals correspond to the {\em large field region}.
In all theses cases  we bound the interaction terms (the trace and $V_j$)
by sup norm in terms of quadratic and linear expressions 
in $a$ and $b$. The large contributions from this bound are then
compensated by the small probability factor (as the large field region
is very unlikely). Note that the $b$ field bounds are more delicate
because of the double well structure  (see Fig.~\ref{inter}).

\vskip 0.2cm

Below we analyze the integration restricted to each interval.

\subsubsection{Small field region: leading contribution $T_1$}

We consider the leading contribution $T_1$.
In the region $I^1$ all the $a$ fields and the $b$ fields  are near 0. 
We apply
\begin{align}
{\rm Re}\, V(a_j)\  &\leq  \ K\    |a_j|^3 
\quad 
{\rm if}\ |a_j|<<1 \no\\
{\rm Re}\, V(b_j)\  &\leq  \ K\    |b_j|^3 
\quad 
{\rm if}\ |b_j|<<1  
\end{align}
and  we bounded the Tr$(D+D'_0)B$ applying 
\begin{align}
|D_{j}|\ &\lesssim  \   |a_j|+|b_j| \\
|D'_{0}|\ &\lesssim  \   1 \no
\end{align}
Therefore we can write 
\begin{equation}
T_1 \leq \ 
\int d\mu_C(a,b)\, e^{ K\sum_j \lp |a_j|^3 + |b_j|^3 \rp}  
e^{K\sum_j \lp |a_j|+|b_j|\rp W^{-2}} \chi[I^1] \no
\end{equation}

Now we insert the cubic and linear contributions
in the measure by this definition
\begin{align}
H(a) &=: a^TC^{-1}a\, - \,\left[ K\,\sum_j \lp |a_j|^3 + 
\frac{|a_j|}{W^2} \rp\right] \label{Hmes1}\\
Z(H)  &=:   \int  da\, e^{-\frac{1}{2}H(a)} \chi[I^1]  \label{Hmes2}\\
d\mu_{H}(a) &=:  \frac{1}{Z(H)} da\, e^{-\frac{1}{2}H(a)}  \chi[I^1] \label{Hmes} 
\end{align}
The same definitions hold for the $b$ fields. Now we have   
\begin{equation}
\int d\mu_C(a,b)\, e^{ K\, \sum_j \lp |a_j|^3 + |b_j|^3 \rp}  
e^{K \,\sum_j \lp |a_j|+|b_j|\rp W^{-2}} \chi[I^1] 
\ = \  \lp\frac{Z(H)}{Z_0}\rp^2
\end{equation}
where the free partition function is
\begin{equation}
Z_0 = \int da\,  e^{-\frac{1}{2} {a}^T C^{-1}a } \ .
\label{Z0}\end{equation}
Therefore we actually have to estimate the normalization factor of the 
interacting measure
This is done through  Lemma~5 below.

\begin{lemma}
With $Z(H)$ and $Z_0$ defined by (\ref{Hmes2}) and (\ref{Z0})
we have 
\begin{equation}
Z(H)\  \leq\ e^{O\lp \frac{|\La|}{W^3}\rp}\  Z_0 
\end{equation}
\end{lemma}

\paragraph{Proof}
Let $H(t)$ defined as
\begin{equation}
H(t)(a) =: a^TC^{-1}a -\left[ K\,  \sum_j t\, \lp |a_j|^3 +
\frac{|a_j|}{W^2}\rp \right] 
\end{equation}
interpolate between $H$ and $C^{-1}$. 
Note that on $I_1$ 
\begin{align}
\ln \left [\frac{Z(H(1))}{Z(H(0))}\right] &= 
 \int_0^1 dt\,  \frac{d}{dt} \ln Z(H(t)) \no\\
  &=  \sum_j \int_0^1 dt\, \int d\mu_{H(t)}(a) \left[  
|a_j|^3 +\frac{|a_j|}{W^2}\right]\no\\
 &\leq   \sum_j \int_0^1 dt\, \int d\mu_{C_f}(a) \left[  
|a_j|^3 +\frac{|a_j|}{W^2}\right] \leq K \frac{|\La|}{W^3}
\label{normH}
\end{align}
where we defined $ d\mu_{H(t)}(a)$ as in (\ref{Hmes}). In the last 
line we used Brascamp-Lieb (\ref{bl1}) together with
\be
H'' \geq   C^{-1} - f m_r^2 = C_f^{-1} > 0
\label{Cf}\end{equation}
which is valid on $I_1$ and for $f=O\lp W^{-\frac{1}{8}}\rp$. 
In general we will use this definition of $C_f$ for  
$f$ a constant   $0<f<1$ or,  when $\La$ is a set of cubes,
$f$ a diagonal matrix  constant on each cube.
Now
\begin{equation}
Z(H(0))= \left[\int d\mu_{C}\  \chi[I^1]\right]\ \leq Z_0
\end{equation}
This ends the proof.
\qed
\vskip 0.2cm

Applying Lemma~5 we have
\begin{equation}
T_1 \ \leq\    e^{O\lp\frac{|\La|}{W^3}\rp} \leq K
\end{equation}

\subsubsection{Small field region: contribution from the second saddle}

In this section we show that $T_2\leq e^{-cW}$. This means 
that the fields have actually the same behavior as in a large field
region. Note that this property holds only in three dimensions.

In the interval $I^2$ all the $a$ fields are near 0
and the $b$ fields are near the second saddle $2{\cal E}_i$.
Recall that ${\cal E}_i=\sqrt{1-E^2/4}$.
Note that  for all $|a_j|\leq   W^{-\frac{1}{8}}$ and
$|b_j-2{\cal E}_i|\leq W^{-\frac{1}{8}}$ we have
\begin{align}
{\rm Re}\, V(a_j) \ \leq & \ \frac{m_r^2}{2}\, f_a a_j^2
\label{b4} \\
{\rm Re}\, V(b_j)\  \leq & \ 
\frac{m_r^2}{2}\, f_b b_j^2 + (1-f_b) m_r^2 2{\cal E}_i \lp b_j-{\cal E}_i\rp 
  \label{b7} 
\end{align}
with $f_a=f_b= O(W^{-\frac{1}{8}})$. 
Note that for $b$ there is a  linear contribution 
coming from the translation to the second saddle. 
Moreover  can be bounded Tr$(D+D'_0)B$ applying Lemma~6 below.

\begin{lemma} 

If $|a_j|\leq W^{-\frac{1}{8}} $ and $|b_j-2{\cal E}_i|\leq W^{-\frac{1}{8}}$, 
 then the real part of $[(D+D'_0)B]_{jj}$  is bounded by
\begin{equation}
{\rm Re}\, (D+D'_0)_j B_{jj} \leq  
 -c\  W^{-2}  
\label{trb}\end{equation}
where $c>0$ is some constant independent from $W$.
\end{lemma}

\paragraph{Proof} 

Note that
\begin{equation}
{\rm Re}\ (DB)_{jj}  =
\left [
{\rm Re}\, D_j \ {\rm Re} \, B_{jj} - {\rm Im}\, D_j\ {\rm Im} \, B_{jj}
\right ] \label{reim} 
\end{equation}
The key point is that, for $a_j$ near zero and $b_j$ near to the second
saddle we have
\begin{align}
{\rm Re}\, D_j = & - m_r^2 + O\lp \frac{1}{W^{\frac{1}{8}}}\rp \no\\
{\rm Im} \,D_j = & - m^2_i  +  O\lp \frac{1}{W^{\frac{1}{8}}}\rp 
\end{align}
Note that this estimates are not true in other regions. If both $a_j$
and $b_j$ are near zero $D_j\simeq 0$ while for $a_j$ or $b_j$ 
far from the saddle we can only say that $|D_j|\leq const$. 
For $D'_0$ we only need to know that for any $a_0, b_0\in \RR$
\begin{equation}
 D'_0 \,=\, O(1) 
\end{equation}
Now, by simple Fourier space analysis we see that
\begin{align}
{\rm Re}\, B_{jj} &\geq \, c \frac{1}{W^2} \\
{\rm Im}\, B_{jj} & =  O\lp \frac{1}{W^3}\rp 
\end{align}
Inserting these estimates in (\ref{reim}) the proof follows.
\qed
\vskip 0.2cm

Inserting all this results in $T_2$ we have
\be
T_2 \leq  e^{-c \frac{|\La|}{W^2}}
\int d\mu_C(a,b)\, e^{\frac{m_r^2}{2}\left[ f_a\sum_j a_j^2 + f_b 
\sum_j b_j^2\right]}  e^{ (1-f_b) m_r^2 2{\cal E}_i 
\sum_j \lp b_j-{\cal E}_i\rp }
\end{equation}

We insert the quadratic terms in the measure.
The normalization ratio are bounded using Lemma~\ref{ratios} below

\begin{lemma} 
For any $0<f<1$ we have
\be
\det \lp\frac{C^{-1}}{C^{-1}_f}\rp \  \lesssim  \ \frac{1}{1-f}\ 
e^{O\lp \frac{f|\La|}{W^2}\rp}
\end{equation}
where $C_f$ is defined in (\ref{Cf}).
\label{ratios}
\end{lemma}

\paragraph{Proof}
Diagonalizing the matrices we can write
\begin{align}
\det \lp\frac{C^{-1}}{C^{-1}_f}\rp &= 
\prod_k \frac{2\sum_{i=1}^3(1-\cos k_i) W^2+
m_r^2}{2\sum_{i=1}^3(1-\cos k_i)W^2+m_r^2(1-f)} \\
 & \leq 
 \frac{1}{1-f} \left |e^{fm_r^2{\rm Tr} \, C^0_{f}}\right|  
 \leq \frac{1}{1-f}\ e^{O\lp \frac{f|\La|}{W^2}\rp}\no
\end{align}
where we defined $C^0_f$ as the covariance $C_f$ where the zero mode
has been extracted.
This ends the proof.\qed \vskip 0.2cm

Therefore we can write
\begin{align}
T_2 &\leq   
\tfrac{1}{(1-f_a)(1-f_b)}  
e^{O\lp \frac{f_a|\La|}{W^2}+\frac{f_b|\La|}{W^2}\rp} 
e^{-c\, \frac{|\La|}{W^2}}
 \int d\mu_{C_{f_b}}(b)  \, e^
{ (1-f_b) m_r^2 2{\cal E}_i \sum_j \lp b_j-{\cal E}_i\rp }  \no\\
&  \leq e^{O(W W^{-\frac{1}{8}}) }\   
e^{-c\,W} \lesssim \ e^{-c\,W} 
\end{align}
where we inserted $|\La|=W^3$ and $f_a=f_b=O(W^{-\frac{1}{8}})$ and
we applied
\begin{equation}
 \int d\mu_{C_{f_b}}(b)  \, 
e^{ (1-f_b) m_r^2 2{\cal E}_i \sum_j \lp b_j-{\cal E}_i\rp }  = 1 \ .
\end{equation}

\subsubsection{Large field region}
This is the region selected by the intervals $I^3$ (one $a$ fields large)
$I^4$ (one pair of $b$ fields with $|b_j-b_{j'}|$ large) and
$I^5$ (all $b$ fields far from both saddles).
We apply the following inequalities
\begin{align}
{\rm Re}\, (D+D'_0)_i B_{ii} \ \leq & \ \sup_{a,b} |(D+D'_0)_i| \ 
 |B_{ii}| \lesssim O\lp \frac{1}{W^2}\rp \label{b1} \\
{\rm Re}\, V(a_j) \ \leq & \ \frac{m_r^2}{2}\, f_a a_j^2 
\label{b2} \\
{\rm Re}\, V(b_j)\  \leq & \ \frac{m_r^2}{2}\, f_b b_j^2 + O(1-f_b) 
\label{b3} 
\end{align}
with $1/2< f_a< 1 $,  $f_b=1-W^{-3}$. 
These estimates are true for any value of $a_j$ and $b_j\in \RR$. 
On the other hand,
when we are in the interval $I^5$, all $b$ fields must be far from both
saddles the  interaction in exponentially small, 
therefore we gain an additional small factor:
\begin{align}
{\rm Re}\, V(b_j)_{|I^5}\  \leq & \ \frac{m_r^2}{2}\, f_b b_j^2 + O(1-f_b)
- c \lp\frac{1}{W^{\frac{1}{8}}}\rp^2 
\label{b31} 
\end{align}
Note that the factor $ O(1-f_b)$ comes from the contribution
of the second saddle (see Fig.~\ref{inter}). 
Therefore we can write
\begin{align}
& T_3+T_4+T_5  \ \leq \ e^{O\lp\frac{|\La|}{W^2}+(1-f_b)|\La|\rp} \cdot \\
&\ \  \cdot \int d\mu_C(a,b)\, 
e^{\frac{m_r^2}{2}\left[ f_a\sum_j a_j^2 + f_b \sum_j b_j^2\right]}
\left \{ \chi[I^3] +  \chi[I^4] + e^{-c\, W^3 W^{-\frac{1}{4}} }
\right\}
\no\end{align}
We insert the quadratic terms in the measure:
\begin{align}
&  T_3+T_4+T_5 \ \leq \ 
\frac{1}{(1-f_a)(1-f_b)}\ 
e^{O\lp \frac{(f_a+f_b)|\La|}{W^2}\rp}
e^{O\lp\frac{|\La|}{W^2}+(1-f_b)|\La|\rp} \cdot 
\no\\
&
\cdot  
 \int d\mu_{C_{f_a}}(a)\   d\mu_{C_{f_b}}(b)\ 
 \left \{ \chi[I^3] +  \chi[I^4] + e^{-c\, W^3 W^{-\frac{1}{4}} }
\right\}
\end{align}
where we defined $C_{f_a}$ and $C_{f_b}$ as in (\ref{Cf})
and  we applied  Lemma~\ref{ratios}.

To bound the contributions from $I^3$ and $I^4$ we apply the following
Lemma.

\begin{lemma}
The probability of having one $|a_j|> W^{-\frac{1}{8}}$ or
one pair $|b_j-b_{j'}|> W^{-\frac{1}{8}}$ is exponentially small
\begin{align}
\int d\mu_{C_{a}}(a)\  \chi[I^3] &\lesssim \ W^3 
e^{-c\, W^2 W^{-\frac{1}{4}} }\\
\int  d\mu_{C_{b}}(b)\  \chi[I^4]  &\lesssim \ W^6 
e^{-c\, W^2 W^{-\frac{1}{4}} }
\end{align}
\end{lemma}

\paragraph{Proof}
We consider first the integral for $a$
\begin{align} 
& \int d\mu_{C_{f_a}}(a)\  \chi[I^3]\  \leq\  \sum_j
\int d\mu_{C_{f_a}}(a)\  \ \chi\lp |a_j|>W^{-\frac{1}{8}}\rp\no\\
& \ \  \leq \ 
\sum_j \int d\mu_{C_{f_a}}(a)\ \frac{ \lp  e^{-xa_j} + 
e^{xa_j} \rp }{1+e^{-2\,x\,W^{-\frac{1}{8}}}} 
\ e^{-x\,W^{-\frac{1}{8}}} \no\\
&\ \  \leq\  \sum_j  2 e^{\frac{1}{2} x^2 (C_{f_a})_{jj}} 
e^{-x\,W^{-\frac{1}{8}}} \lesssim  
W^3 e^{-c\,W^2\,W^{-\frac{1}{4}}}
\end{align}
where we applied $(C_{f_a})_{jj}=O(1/W^2)$ and we  
set $x=O( W^{-\frac{1}{8}}W^2)$.
The same proof holds for the $b$ field. In this case the 
presence of a difference
$b_j-b_{j'}$ is crucial to ensure that the factor 
 $[(C_{f_b})_{jj} +(C_{f_b})_{j'j'}-2(C_{f_b})_{jj'}]$ is 
of order  $W^{-2}$ and does not
depend on the mass (which could be very tiny for $C_{f_b}$).  
The factor $W^6$ comes form the sum over $j$ and $j'$.
\qed \vskip 0.2cm

Putting together all the factors we have
\begin{align} 
&  T_3+T_4+T_5 \ \leq \ 
\frac{1}{(1-f_a)(1-f_b)}\ 
e^{O\lp \frac{(f_a+f_b)|\La|}{W^2}\rp}
e^{O\lp\frac{|\La|}{W^2}+(1-f_b)|\La|\rp} \cdot 
\no\\
&
\cdot  
 \left \{ (W^3+W^6)  e^{-c\, W^2 W^{-\frac{1}{4}} } 
 +   e^{-c\, W^3 W^{-\frac{1}{4}} }
\right\}\  \lesssim   e^{-c\, W^3 W^{-\frac{1}{4}} }
\  
\end{align}
where we have inserted $|\La|=O(W^3)$, $f_a=3/4$ and $f_b=1-(1/W^3)$.
Note that there is an additional factor $W^3$ from the zero mode 
$(1-f_b)^{-1}$ of the determinant.

\subsubsection{Sum over the different regions}
Summing the bounds on different intervals we have finally

\be 
\left |\left\langle G^+_{00}\right\rangle_{\vep =0} \right | 
\lesssim  \left[ 1 + e^{-c\,W} + e^{-c\,W^{3-\frac{1}{4}}}\right] \leq K
\end{equation} 
This completes the proof of (\ref{boundrho1}).

\subsubsection{Large volume}

It is straightforward to extend the above estimates to the
case when $\La$ is a union of cubes.

\begin{coroll}
The density of states in a union of cubes $\La$ is bounded by 
\be
\left | {\bar\rho}_\La(E)  \right | \lesssim \  e^{O\lp\frac{|\La|}{W^3}\rp}
\label{boundrho2}\end{equation}
\end{coroll}

\paragraph{Proof}

In each cube we apply the bounds above. The result is written as a quadratic form
$\exp[v^T C_f v]$ where $v$ is a vector which depends on the
bounds on each particular cube and $f$ is now a diagonal matrix which
is constant on each cube.
The key point is that 
\be
 C_f \leq \frac{1}{- W^2 \De_N + (1-f) m_r^2}
\end{equation} 
where $\De_N$ is a Laplacian with Neumann boundary conditions
on the cubes, and decouples the cubes automatically. 
Now we can perform the estimates in each cube separately.
This completes the proof of (\ref{boundrho2}).
\qed \vskip 0.2cm

\subsection{Proof of Theorem 3}

To prove this result  we integrate by parts to generate
perturbative terms. To control the remainder we apply the bounds of 
Theorem~2.

\subsubsection{Semicircle law}

We prove that ${\bar\rho}(E)=\rho_{SC}$ with a precision of order $W^{-2}$: 
\begin{equation}
{\bar\rho}_\La(E) =\rho_{SC} + O\lp\frac{1}{W^2}\rp \ .
\label{semi}\end{equation}
Note that ${\bar\rho}_\La(E) =-\frac{1}{\pi}{\rm Im} \langle G^+_{00}\rangle$ 
therefore we have to study
\begin{equation}
\langle G^+_{00}\rangle = \int d\mu_B(a,b)\  \frac{1}{({\cal E}^*-a_0)}  
e^{V} \det[1+(D+D'_0)B]
\label{expres}\end{equation}
We have to perform a few steps of perturbative expansion on the
observable $({\cal E}^*-a_0)^{-1}$ and $D'_0$. These are more clear if we 
write the determinant as a fermionic integral.
\begin{equation}
 \det[1+(D+D'_0)B] = 
 \int d\mu_B(\rho^*,\rho) \   e^{-\rho^*\rho D} \lp 1 - D'_0 \rho^*_0\rho_0\rp
\end{equation}
where we defined
\begin{equation}
d\mu_B(\rho^*,\rho) \  = \  \det B \  e^{-\rho^* B^{-1}\rho}
\end{equation}
\begin{equation}
\rho^*\rho D \  = \ \sum_{j} \rho^*_j\rho_j D_j 
\end{equation}
and $D_j$ and $D'_0$ are introduced in (\ref{vertex2}) and  (\ref{vertex4}).
The density of states is then written as
\begin{equation}
  \langle G^+_{00}\rangle = \int d\mu_B(a,b,\rho^*,\rho) \ 
  e^{V-\rho^*\rho D}  \ {\cal O}_0
\label{fermexpr}\end{equation}
where we defined  $ d\mu_B(a,b,\rho^*,\rho) =d\mu_B(a,b)d\mu_B(\rho^*,\rho)$
and   the observable $ {\cal O}_0$ is
\begin{align}
 {\cal O}_0 & =  \frac{1}{({\cal E}^*-a_0)} \lp 1 - D'_0 \rho^*_0\rho_0\rp
\label{observ}\\
          & = {\cal E} + a_0 \int_0^1 dt  \frac{1}{({\cal E}^*-t a_0)^2} 
           -  \frac{1}{({\cal E}^*-a_0)} D'_0 \rho^*_0\rho_0 \no
\end{align}
The first term is a constant and gives the semicircle law 
$-\tfrac{1}{\pi}{\rm Im}  {\cal E}= \rho_{SC}$. Note that we apply
\begin{equation}
 \int d\mu_B(a,b,\rho^*,\rho) \ 
  e^{V-\rho^*\rho D}  \ =\ 1
\end{equation}
The remaining two terms give the corrections
\begin{align}
\de\rho_1 &= \int d\mu_B(a,b,\rho^*,\rho) \ e^{V-\rho^*\rho D}  
\left [a_0 \int_0^1 dt  \frac{1}{({\cal E}^*-t a_0)^2} \right] \\
\de\rho_2 &= \int d\mu_B(a,b,\rho^*,\rho) \  e^{V-\rho^*\rho D}  
\left [ - D'_0 \rho^*_0\rho_0\right]
\end{align}

\paragraph{Estimate of $\de\rho_2$}
We first consider the estimate on the second integral, as it is
the easiest one.
We partition the integral domain inserting (\ref{partition})
as in Sec.~5.1 and we perform the fermionic integral in a different
way depending on the region.

  Near the first saddle (interval $I^1$)  we
apply
\begin{equation}
\int d\mu_B(\rho^*,\rho) \  e^{-\rho^*\rho D}    \  \rho^*_0\rho_0
\ =\ \lp \frac{1}{B^{-1}+D}\rp_{00} \det[1+DB]  
\label{roint1}\end{equation}
It is easy to see that for $a_j$ and $b_j$ near zero $D_j\simeq 0$
and $|(B^{-1}+D)^{-1}_{00}|= O(W^{-2})$.
Therefore we have
\begin{equation}
|\de\rho_2(I^1)| \lesssim \frac{1}{W^2}  \int \left |  d\mu_B(a,b)\ 
 e^{V} \det[1+DB]  
\left [ \frac{-D'_0}{({\cal E}^*-a_0)} \right]  \right| \ \lesssim  \frac{1}{W^2}
\end{equation}
where we applied the same bounds as in Sec.~5.1.1.

 In the other regions $I^k$, $k\neq 0$ we cannot apply 
(\ref{roint1}) as $(B^{-1}+D)^{-1}$ is not well defined ($D$ is big
and may cancel  $B^{-1}$). Therefore we apply
\begin{equation}
\int d\mu_B(\rho^*,\rho)\   e^{-\rho^*\rho D}    \  \rho^*_0\rho_0
\ =\  \det\, M 
\end{equation}
where $M$ is the matrix $1+DB$ with  the row $0$
 substituted with $B$:
\begin{equation}
\ba{ccccc}
M_{ij} &=& (1+DB)_{ij} &\quad &  i,j\neq 0 \\
M_{i0} &=& D_iB_{i0} &\quad &  i\neq 0 \\
M_{0j} &=& B_{0j} &\quad &  \forall \, j \\
\ea
\end{equation}
If we apply (\ref{detb}) we obtain the same bounds as in Sec.~5.1.
Therefore performing the same bounds as in  Sec.~5.1.2-5.1.3,
we have
\begin{equation}
\sum_{k\neq 1}|\de\rho_2(I^k)| \lesssim \ e^{-c\, W}
\end{equation}
Hence 
\begin{equation}
|\de\rho_2| \lesssim \  \frac{1}{W^2} + e^{-c\, W} \lesssim \  \frac{1}{W^2} 
\end{equation}

\paragraph{Estimate of $\de\rho_1$} Now we consider the first error term.
Before inserting the partition (\ref{partition}) and integrating
over the fermionic integrals we have to perform one step of integration by
parts
\begin{align}
\de\rho_1 &= \sum_k B_{0k} \int d\mu_B(a,b,\rho^*,\rho) \ e^{V-\rho^*\rho D}  
\\ 
& \qquad \times \left [
\frac{\de}{\de a_k} \lp V_k - \rho^*_k\rho_k D_k\rp + \de_{0k} 
\frac{\de}{\de a_0} \int_0^1 dt  \frac{1}{({\cal E}^*-t a_0)^2} \right] \no
\end{align}
Note that $|\tfrac{\de}{\de a_k}  V_k|\lesssim |a_k|^2 +  |a_k|^3$. 
In the region around the first saddle  ($I^1$),  
applying the Brascamp-Lieb inequality (\ref{bl1}),
these fields give a factor $W^{-2}$. In the other regions they are 
bounded by the exponential mass decay. The contribution from $\rho^*_k\rho_k$
is estimated as in $\de\rho_2$ above.
Therefore
\begin{equation}
|\de\rho_1| \lesssim \  \lp\frac{1}{W^2}\sum_k |B_{0k}|\rp + |B_{00}| \lesssim \  
\frac{1}{W^2} 
\end{equation}
where we applied $\sum_k  |B_{0k}|\leq const$ and  $ |B_{00}|= O(W^{-2})$.
This ends the proof of (\ref{semi})

\subsubsection{Smoothness}
Now we consider the derivatives. Note that it is easier to compute
the derivatives  on the starting expression 
$ \langle G^+_{00}\rangle$  than directly on the functional integral (\ref{expres}).
The derivative at order $n$ is given by
\be
\partial_E^n {\bar\rho}_\La(E) = - n! (-1)^n \tfrac{1}{\pi}{\rm Im}
 \left\langle (G^n)_{00} \right \rangle \propto
 {\rm Im} \sum_{j_1,...,j_n}  
\left\langle G_{0j_1}...G_{j_n0} \right \rangle
\end{equation}
Applying the supersymmetric approach and the saddle point 
analysis as in Sec.~3-4, we can write for instance $R(x)$ as
\begin{equation}
R(x) = 
\left\langle G_{0x} G_{x0} \right \rangle
= \left\langle {\cal O}_0\   {\cal O}_x  \right \rangle_{SUSY} -
 \left\langle {\cal O}_0   \right \rangle_{SUSY}
 \left\langle  {\cal O}_x  \right \rangle_{SUSY}
\end{equation}
where we defined 
\begin{equation}
\left \langle F(a,b,\rho^*,\rho)  \right \rangle_{SUSY} 
=:   \int d\mu_B(a,b,\rho^*,\rho) \  e^{V-\rho^*\rho D}  F(a,b,\rho^*,\rho)
\end{equation}
and the observables are (\ref{observ}) and 
\begin{equation}
 {\cal O}_x   =  \frac{1}{({\cal E}^*-a_x)} \lp 1 - D'_x \rho^*_x\rho_x\rp\ .
\end{equation}
A similar formula holds for the general case.

We perform now integration by parts starting from ${\cal O}_0$
until we have a path of connected vertices that connects 0 to $j$ or
 we have enough vertices to extract a factor $W^{-3}$ for each observable
${\cal O}_j$. This factor ensures that we can sum over the position of $j$ inside the
cube $\La$.

\noindent Note that, as in general we will have to estimate
products of fields, both fermionic and bosonic, we will
need the two Lemmas below.

\begin{lemma}
Let consider the average of the product of  $p$ fermionic fields 
\be
\int d\mu_B(\rho^*,\rho)\  e^{-\rho^*\rho D} 
\left [\prod_{k=1}^{p} \rho_{i_k}  \rho^*_{j_k} \right]
\end{equation}
Note that $i_k$ and $j_k$ are not necessarily equal.
This integral gives different estimates depending on the
region we are considering. If $a$ and $b$ are near zero we have
\be
\left |\int d\mu_B(\rho^*,\rho)\  e^{-\rho^*\rho D}  \chi[I^1]
\left [\prod_{k=1}^{p} \rho^*_{i_k}  \rho_{j_k} \right]\right |\ 
\leq \ \frac{p!^2 }{W^{2p}}  \ |\det(1+DB)|\  \chi[I^1]
\label{fbound}\end{equation}
On the other hand, in the other regions $I^s$,  $s\neq 1$ we have
\be
\int d\mu_B(\rho^*,\rho)\  e^{-\rho^*\rho D}  \chi[I^s]
\left [\prod_{k=1}^{p} \rho^*_{j_k}  \rho_{i_k} \right]\ 
= \  \si \ \left[{\det}_{JI} M\right] \chi[I^s]
\label{sbound}\end{equation}
where $M$ is the matrix $1+DB$ where the rows $i_1$,...$i_p$
are substituted by the corresponding rows of $B$, and
the columns  $j_1$,...$j_p$  are substituted by 
the corresponding columns of $DB$:
\begin{equation}
\ba{ccccc}
M_{ij} &=& (1+DB)_{ij} &\quad &  i\neq i_1,...i_p,\, j\neq j_1,...j_p \\
M_{ij_k} &=& D_iB_{ij_k} &\quad &  i\neq i_1,...i_p \\
M_{i_kj} &=& B_{i_kj} &\quad &  \forall \, j \\
\ea
\end{equation}
Finally  $I$ are the set of indices $I=\{i_1,...,i_p\}$ $J=\{j_1,...,j_p\}$
and $\det_{JI}M$ is the determinant of the matrix 
$M$ without the rows $j_1,,...j_p$ and the columns 
$i_1,...,i_p$, and $\si$ is a sign. 
This new determinant can be bounded as usual. 
\be
| {\det M}_{JI}| \lesssim e^{O\lp \frac{|\La|}{W^3}\rp}
 e^p\left| e^{\sum_{i\not\in I\cup J} (DB)_{ii}}\right| 
\label{sbound1}\end{equation}
\end{lemma}

\paragraph{Proof}
To obtain the first bound (\ref{fbound}) we apply (\ref{roint1}) for a product of fermionic
fields. The result is the determinant of a $p\times p$ matrix whose elements
are $(B^{-1}+D)^{-1}_{ij}$ with $i=i_1,...,i_p$ and $j=j_1,...j_p$.
This determinant is easily bounded by $p!^2 \sup |(B^{-1}+D)^{-1}_{ij}|$.
Applying $ |(B^{-1}+D)^{-1}_{ij}|= O(W^{-2})$ we obtain the result.

The second expression (\ref{sbound})
is easily obtained using the anticommuting properties of the
fermionic fields.
Finally  (\ref{sbound1}) holds because the only error terms
come from the absence of a term 1 in p diagonal elements.
\qed \vskip 0.2cm

\begin{lemma}
We consider the integral
\begin{equation}
I_a(n_1,...,n_p)\ =:\  \int d\mu_C(a,b) \  \left | e^{V} \right|
\  \prod_{k=1}^p\  |a_{j_k}|^{n_k} \ e^{{\rm Re}\, {\rm Tr}\, DB}
\end{equation}
where $p>0$, $n_k>0$ for all $k$ and $n=\sum_k n_k$.  Then 
\begin{align}
I_a(n_1,...,n_p)[I^1] \ & \lesssim\  n!\, \lp\frac{K}{W}\rp^n \  \label{sfb1}\\
I_a(n_1,...,n_p)[I^2] \ &\lesssim\ K^n \ 
 e^{-c W} \label{sfb2}\\
I_a(n_1,...,n_p)[I^q] \ &\lesssim\ K^n \ \prod_k\sqrt{n_k!}\ 
 e^{-c W^{\frac{7}{4}} }\qquad q>2 \label{sfbk}
\end{align}
If instead of $a$ fields we have $b$ fields the result is the same,
but in the large field region we pay a larger factor, because
we have a very small mass remaining in the covariance
\be
I_b(n_1,...,n_p)[I^q] \ \lesssim\ K^n W^{\frac{3}{2}n}\  \prod_k\sqrt{n_k!}\ 
 e^{-c W^{\frac{7}{4}} }\qquad q>2 \label{sfbkb}
\end{equation}
\end{lemma}

\paragraph{Proof}
As in Sec. 5 we partition the integration domain 
$1=\sum_{q=1}^6 \chi[I^q]$.

When we are near the first saddle ( $I^1$) we write
\be
|a_{j_1}|^{n_1}\ ...\ |a_{j_p}|^{n_p}\  \leq\ 
 \frac{1}{n}\  \sum_{k=1}^p\ n_k\  |a_{j_k}|^n
\end{equation}
Now we can apply Brascamp-Lieb inequality as stated in (\ref{bl1}).

In the region near the second saddle  ( $I^2$) we can bound the
field $a$ by a constant. 

In the large field region  we bound the fields $a$ using a fraction
of the exponential decay of the mass term 
\be
|a_j|^n\ \leq\ \lp\frac{K}{\sqrt{\de}}\rp^n \ \sqrt{n!}\  
e^{\frac{1}{2}\de m_r^2 a_j^2}
\end{equation}
where $\de>0$ is a small constant $\de<1$ which must be smaller
than the mass of $C_{f_a}$. 
Note that, for the $b$ fields in the region $I^3$ or $I^4$ 
$\de$ must be of order $\de=O(W^{-3})$ as this is the mass of   $C_{f_b}$.
This completes the proof.
\qed \vskip 0.3cm

\section{Infinite volume limit}
\resetequ

In this section we shall establish bounds on $\langle G^+_{00}\rangle$
and the exponential decay of $R(x)$ uniformly as  $\La\uparrow \ZZZ^3$.
This is done by a standard method (see \cite{Bry} or \cite{R}, ch.III.1) 
in statistical mechanics called the cluster expansion.  
These expansions are possible when there is a single dominant saddle point 
(in our case $a=b=0$) whose fluctuations are close to that of a massive 
Gaussian i.e. a Gaussian whose covariance $B$ has exponential decay.
We are going to use a standard expansion with a few modifications.
By supersymmetry some terms of the expansion are one 
(see Lemma~11) thus simplifying the expression. On the other hand
for technical reasons the treatment of the covariance in the measure
is slightly different form the usual one.

We prove the following theorem
\begin{theor} 
There exists $W_0$ such that for all $W>W_0$  
$\lim_{\La\uparrow\ZZZ^3}{\bar\rho}_\La(E)$ 
is bounded in ${\cal I}$ uniformly in $W$
\be
\left |\lim_{\La \uparrow \ZZZ^3} {\bar\rho}_\La(E)  \right | \leq K
\label{rhob}\end{equation}
for some constant  independent from $W$.
Moreover  
\begin{align}
\left |\lim_{\La\uparrow\ZZZ^3}\partial_E^n {\bar\rho}_\La(E)  \right |& \leq   \ 
C_n \qquad \forall\, n<n_0(W)\label{derivn}\\
\lim_{\La\uparrow\ZZZ^3} {\bar\rho}_\La(E)  & 
= \rho_{SC}(E) +  O\lp W^{-2}\rp \label{semic}
\end{align}
uniformly in $\La$ and $W$.
In particular, for $x\neq 0$
\be
\left |\lim_{\La\uparrow\ZZZ^3} R(x)|  \right | \leq \frac{K}{W^3}\
e^{-c \frac{|x|}{W}}
\label{Rdec}\end{equation}
\end{theor}

\paragraph{Outline of the proof:}
Note that the exponential decay (\ref{Bdecay})  of $B$ means that
 regions at a distance higher than $W$ 
are approximately decoupled. 
As the observable depends only on the fields at zero or at zero and $x$ 
we expect that all  interactions take place
in a volume of order $W^3$ around $i=0$.  
To exploit this fact we   partition $\La$ in cubes of side
$W$  forming the lattice ${\cal L}$. 
For that we introduce the function
\be
\chi_\De(i)  =
\begin{cases}
 1  & {\rm if} \quad i\in \De \\
 0  & {\rm otherwise} 
\end{cases}
\end{equation}
satisfying $\sum_{\De\in {\cal L}}\chi_\De(i)= \chi_\La(i) $. 
In the following we call {\em root cube} the cube containing $i=0$
and we denote it by $\De_0$. 

The cluster expansion expresses $\langle G_{00}^+\rangle$ and $R(x)$
as a sum of finite volume contributions. 
Let $Y$ be a union of cubes $\De$ containing zero (zero and $x$ when we 
estimate $R(x)$). Then the cluster expansion gives 
\be
\left\langle G^+_{00}\right\rangle_{\vep =0} = \sum_{Y\ni \De_0} c_Y Z'_Y 
\label{clust}\end{equation}
where $Z'_Y$ is a functional integral over fields $a_j$, $b_j$, $\rho_j$ and 
$\rho^*_j$, $j\in Y$, and  $c_Y$ is a coefficient which is exponentially small
in the length of the shortest tree spanning the cubes of $Y$. The main 
purpose of this section is to give a precise description of (\ref{clust}) and 
provide estimates to establish the convergence of the sum. Note that 
in conventional statistical mechanics there is usually 
 an additional factor $Z_{Y^c}$ with  $Y^c=\La\backslash Y$.
However we show in Lemma~11 below that $Z_{Y^c}=1$.

The factors  $Z'_Y$  are similar to a partition function except that
there are derived vertices $a_j^3$, $b_j^3$ or $\rho^*_j\rho_j(a_j+b_j)$ present.
We shall show, using the ideas of Sec.5.1.5, that $Z'_Y\leq (K W^{-\frac{1}{6}})^{|Y|}$.

\begin{lemma} If we restrict to the set of cubes $Y^c=\La\backslash Y$
and there is no observable contribution we have
\be
Z_{Y^c}  =\int d\mu_{B_{Y^c}}(a,b,\rho^*,\rho)  \  
e^{\sum_{j\in \La'}(V_j-\rho^*_j\rho_j D_j)} = 1
\end{equation} 
where $ d\mu_{B_{Y^c}}(a,b,\rho^*,\rho)$ is defined after (\ref{fermexpr})
and $B_{Y^c}$ is the covariance $B$ restricted to the volume $Y^c$.
\end{lemma}

\paragraph{Proof} 
We perform the translation $a_j\rightarrow a_j-{\cal E}$, 
 $b_j\rightarrow b_j+i {\cal E}$ for all $j\in \La'$. Note that,
for a general $\La'\subset \La$, 
$B^{-1}_{\La'}\neq -W^2\De + (1- {\cal E}^2)$. Therefore the translation
gives some linear and constant terms. The constant terms are cancelled
when we add the contributions from the $a$ and $b$ fields.
By performing the inverse Hubbard-Stratonovich transformation
we obtain
\be
R_{\La'}  = \left \langle \int d\Phi^* d\Phi\ 
e^{i \Phi^+ (E_\vep+A-H)\Phi} \right \rangle_1 = 1 
\end{equation} 
where the average $\langle  \  \rangle_1$ is
computed with the probability distribution (\ref{proba}) with covariance  
${\tilde J}$ instead of  $J$,  with ${\tilde J_{ij}}= B^{-1}+{\cal E}^2$.
The matrix $A$ is a diagonal matrix 
\be
\Phi^+ A\Phi = \sum_i  \Phi^+_i\Phi_i A_i\ , \qquad 
A_i= {\cal E} \sum_k ({\tilde J}_{ik} - \de_{ik})
\end{equation} 
This completes the proof.
\qed \vskip 0.2cm

\subsection{Cluster expansion}

We derive the cluster expansion formula. 
The result is stated in Lemma~12 below.
We construct the expansion  by an inductive argument.

First we want to test if there is any connection between 
the root cube $\De_0$ and some other $\De\in \La$.
For that purpose we introduce an interpolated covariance $B(s_1)$ 
with $0\leq s_1\leq 1$, 
which satisfies $B(1)=B$ while $B(0)$ decouples the root cube
$\De_0$  from the rest of the volume.
The easiest choice for $B(s)$ is $B(s)_{ij}=s B_{ij}$ for $i\in \De_0$  
and $j\not\in\La\backslash \De_0$, or vice versa, and  $B(s)_{ij}= B_{ij}$
otherwise. For technical reasons we choose the
following (less natural) interpolation rule
\be
B(s_1)^{-1} = C(s_1)^{-1} + i  m_i^2 
\label{bs}\end{equation}
where
\be
C(s_1)_{ij} =
\begin{cases} 
s_1 C_{ij}  & {\rm if} \quad
i\in \De_0, j\in \De\neq \De_0,  \quad {\rm or\  vice \ versa}  \\
\ \  C_{ij}   &  {\rm otherwise} \ .\\
\end{cases} 
\label{interpol1}
\end{equation}
The reason we use this definition of $B(s)$ 
is that we do not want to mix the real and imaginary part in $B^{-1}$
in order  to apply later the same estimates of Sec.~5.
Note that (\ref{interpol1}) is equivalent to the definition
\be
C(s_1) = s_1 C + (1-s_1) 
\left [ C_{\De_0\De_0}  +  
 C_{ \De_0^c\De_0^c}\right ]
\end{equation}
\be
(C_{\De\De'})_{ij}= \frac{1}{2} \left [
\chi_{\De}(i) C_{ij} \chi_{\De'}(j) + 
\chi_{\De'}(i) C_{ij} \chi_{\De}(j)
\right ]
\end{equation}
where  $ \De_0^c = {\cal L}\backslash \De_0$. 
Therefore  $C(s)$ is still  a positive operator, as it
is a convex combination of the positive operators $C$
and $C_{\De\De}$. This fact is essential to ensure
the convergence of the integrals.
With the interpolated covariance we define 
\be
F_\La[s_1] = \int d\mu_{B(s_1)}(a,b,\rho^*,\rho)  \  
e^{\sum_{j\in Y^c}(V_j-\rho^*_j\rho_j D_j)} \ {\cal O}_0 \ .
\end{equation}
Note that  for $s_1=1$ $F_\La[s_1]_{s_1=1} =
\left\langle G^+_{00}\right\rangle_{\vep =0} $.
Now we apply a first order Taylor formula to $F_\La[s_1]$
\be
F_\La[s_1]_{s_1=1} = F_\La[s_1]_{s_1=0} + \int_0^1 ds_1 \ 
\partial_{s_1} F_\La[s_1]   \ .
\label{taylor}\end{equation}
The first term  $F_\La[s_1]_{s_1=0} = F_{\De_0}$  corresponds to
decoupling $\De_0$ from the rest of the volume.
The derivative in the second term of (\ref{taylor}) gives 
\be
\partial_{s_1} F_\La[s_1] = 
 \int \partial_{ s_1} d\mu_{B(s_1)}(a,b,\rho^*,\rho)  \  
e^{\sum_{j}(V_j-\rho^*_j\rho_j D_j)} \ {\cal O}_0 \ .
\end{equation}
Using integration by parts we have
\begin{align}
\int \partial_{s_1}d\mu_{B(s_1)}(a,b ,\rho^*,\rho)\
&=\int d\mu_{B(s_1)}(a,b,\rho^*,\rho )\ \sum_{ij} 
 \partial_{s_1} B(s_1)_{ij} \ \\
& \times\left [
\frac{\de}{\de a_i}  \frac{\de}{\de a_j}
+  \frac{\de}{\de b_i}  \frac{\de}{\de b_j} +
  \frac{\de}{\de \rho^*_j}  \frac{\de}{\de \rho_i} \right ] \ .
\end{align}
The derivative $\de/\de a_i$ may apply to  $\exp[V_i-\rho^*_i\rho_i D_i]$ 
or to the observable
${\cal O}_0$ (this  only for $i=0$):
\begin{align}
 \frac{\de}{\de a_i} e^{\sum_{j} V_j}   & = e^{\sum_{j} V_j}    
  \left[\frac{\de}{\de a_i} V_i(a_i)\right] \label{Vder}\\
\frac{\de}{\de a_i}  e^{-\sum_{j} \rho^*_j\rho_j D_j}    & =  
\left[\frac{\de}{\de a_i} D_i(a_ib_i)\right] (-\rho^*_i\rho_i) 
e^{-\sum_{j\neq i} \rho^*_j\rho_j D_j}  
\label{detder1} \\
 \frac{\de}{\de a_0} {\cal O}_0  & = (-\rho^*_0\rho_0)  \frac{\de}{\de a_0} D'_0 \ .
\end{align}
The same definitions hold for $\de/\de b_i$.  The fermionic derivative
 $\de/\de \rho_i$  may apply to
$\exp[-\rho^*_i\rho_i D_i]$ or to the observable
${\cal O}_0$ (this only for $i=0$):
\begin{align}
\frac{\de}{\de \rho_i}  e^{-\sum_{j} \rho^*_j\rho_j D_j}    & =  
 D_i \rho^*_i  e^{-\sum_{j\neq i} \rho^*_j\rho_j D_j}  
\label{detder2} \\
\frac{\de}{\de \rho_0} {\cal O}_0  & = \rho^*_0   D'_0 \ .
\end{align}
Similar formulas hold for  $\de/\de \rho^*_i$.
Therefore
\be
 \partial_{s_1} F_\La[s_1]  =  
\sum_{i_1,j_1} \ 
\left [\partial_{s_1} B(s_1)_{i_1j_1}\right ] \ 
F_\La[s_1]({\scriptstyle (i_1j_1)})
\label{taylor1}\end{equation}
where
\begin{align}
F_\La[s_1]({\scriptstyle (i_1j_1)}) &= 
 \int d\mu_{B(s_1)}(a,b,\rho^*,\rho)\ 
 \left [\frac{\de}{\de a_{i_1}} \frac{\de}{\de a_{j_1}} +
\frac{\de}{\de b_{i_1}} \frac{\de}{\de b_{j_1}} + 
 \frac{\de}{\de \rho^*_{j_1}}  \frac{\de}{\de \rho_{i_1}}  
\right ]  \no \\
&\qquad \times \left [e^{\sum_{j}(V_j-\rho^*_j\rho_j D_j)} \ {\cal O}_0 \right ]\ .
\end{align}
Let us consider the propagator $\partial_{s_1} B(s_1)_{i_1j_1}$ extracted
by the Taylor formula. If we choose the easiest interpolating
rule, that is $B(s)_{ij}=sB_{ij}$ when $i$ and $j$ are in different cubes
and $i$ or $j\in\De_0$, the derivative is not zero only for $i_1\in\De_0$
and $j_1\not\in\De_0$, or vice versa. Hence  $\partial_{s_1} B(s_1)_{i_1j_1}$
connects explicitly $\De_0$ to a different cube. 
With the definition (\ref{bs}) the derivation is different and 
instead of one line we extract three
\bqa
\partial_{s_1} B(s_1)_{i_1j_1} &= &\partial_{s_1}
\lp
\frac{1}{C(s_1)^{-1} +i m_i^2} 
\rp_{i_1j_1} \\
&=& \sum_{k_1,k_1'} \lp
\frac{1}{1 +i m_i^2 C(s_1)}\rp_{i_1k_1}  
\left [\partial_{s_1} C(s_1)\right ]_{k_1k_1'}
\lp\frac{1}{1 +i m_i^2 C(s_1)}
\rp_{k_1'j_1} \no\\
&=& \sum_{\De_1\neq \De_0} \sum_{\substack{
k_1\in \De_0 \\ k_1'\in \De_1}}
G(s_1)_{i_1k_1}\,  
C_{k_1k_1'}\,
G(s_1)_{k_1'j_1} \no
\eqa
where we used (\ref{interpol1}) and $G(s)$ is
\be
G(s_1) =: \frac{1}{1 +i m_i^2 C(s_1)} \ .
\label{Gdef1}\end{equation}
For each term $(k_1,k_1')$  with $k_1\in \De_0$ and
$k_1'\in \De_1$ we say there is a {\em strong connection}
between $\De_0$ and $\De_1$. We denote this by drawing a 
line from $\De_0$ to $\De_1$. Note that these points do not
correspond to any derivative inside the functional integral,
as the only derivatives occur on $i_1$ and $j_1$. 
If  $i_1$ and $j_1$ belong to some cube 
 $\De\not\in\De_0\cup\De_1$ they  give some additional strong
connections. 

\begin{figure}
\centerline{\psfig{figure=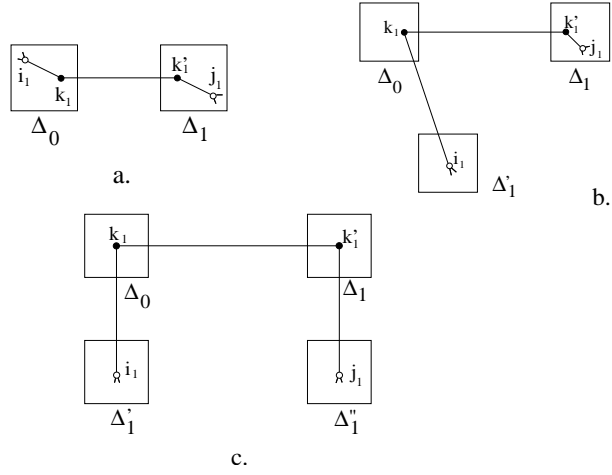,width=8cm}}
\caption{some examples of links of type 1,2 and 3; the two point vertex is  a
filled dot, the three point one is an empty dot}
\label{figlink}
\end{figure}

Therefore the first step of the induction extracts a link 
 $l_1$, associated to four points  $i_1,j_1,k_1,k'_1$, connecting  $\De_0$
to a set of  one, two or three  new
cubes depending on the  positions of $i_1$ and $j_1$. 
We call this set the {\em generalized cube}  ${\tilde\De}_1$.
The different possible links inside ${\tilde\De}_1$ are shown
in Fig.\ref{figlink}.
Now 
\be
 \partial_{s_1} F_\La[s_1]  =
\sum_{\substack{ (i_1,j_1) \\ (k_1,k'_1)}} \ 
G(s_1)_{i_1k_1}  
C_{k_1k_1'}
G(s_1)_{k_1'j_1}
 \ F_\La[s_1]({\scriptstyle (i_1j_1)}) \ .
\label{taylor11}\end{equation}
Note that the functional integral after $\partial_{s_1}B_{s_1}$
has been extracted is function only of $(i_1,j_1)$ and
not of $(k_1,k_1')$, as only the first two indices correspond to
a functional derivative inside the integral.

Now we fix  the points
${(i_1,j_1),(k_1,k'_1)}$ 
corresponding to a strong connection between  $\De_0$ and 
${\tilde\De}_1$.
We want to test if there is any connection between the set  
 ${\tilde \De}_{0,1}=\De_0\cup {\tilde\De}_1$ and any other cube 
$\De'\in \La\backslash ({\tilde \De}_{0,1}) $.
We introduce then a new parameter $s_2$ in $C(s_1)$ in the functional integral:
\be
C(s_1,s_2)_{ij} =
\begin{cases} 
s_2 C(s_1)_{ij}  & {\rm if} \quad
i\in {\tilde \De}_{0,1}, j\not\in {\tilde \De}_{0,1},  \quad {\rm or\  vice \ versa} \\
\ \   C(s_1)_{ij}  &  {\rm otherwise} \ .\\
\end{cases} 
\label{interpol2}
\end{equation}
As for $C(s_1)$ we can write $C(s_1,s_2)$ as a convex combination of positive
operators
\be
C(s_1,s_2) = s_2 C(s_1) + (1-s_2) 
\left [ C_{{\tilde\De}_{0,1} {\tilde\De}_{0,1}}(s_1)  +  
 C_{ {\tilde\De}_{0,1}^c  {\tilde\De}_{0,1}^c}(s_1)\right ]
\end{equation}
where  $ C(s_1)$ and $ C_{\De\De}(s_1)$ are positive. Therefore  
$C(s_1,s_2)$  is still positive.
Then $F_\La[s_1]({\scriptstyle (i_1j_1)})
= F_\La[s_1,s_2]({\scriptstyle (i_1j_1)})_{s_2=1}$.
We apply again the first order Taylor formula
\be
 F_\La[s_1,s_2]({\scriptstyle (i_1j_1)})_{s_2=1} = 
F_\La[s_1,s_2]({\scriptstyle (i_1j_1)})_{s_2=0}
+ \int_0^1 ds_2 \ 
\partial_{s_2}   F_\La[s_1,s_2]({\scriptstyle (i_1j_1)}) \ .
\label{taylor2}\end{equation}
As before $ F_\La[s_1,s_2]({\scriptstyle (i_1j_1)})_{s_2=0}$ 
corresponds to 
a functional integral restricted to ${\tilde\De}_{0,1}$:
 $ F_\La[s_1,s_2]({\scriptstyle (i_1j_1)})_{s_2=0}
= F_{{\tilde\De}_{0,1}}[s_1]({\scriptstyle (i_1j_1)})$.
The other term gives
\begin{align}
\partial_{s_2}   F_\La[s_1,s_2]({\scriptstyle (i_1j_1)}) =
\hspace{-0.2cm} \sum_{\substack{ (i_2,j_2)  (k_2,k'_2) \\ k_2\in{\tilde\De}_{0,1},
 k'_2\in{\tilde\De}_{0,1}^c  }} &  \left[G(s_1,s_2)_{i_2k_2}  
C(s_1)_{k_2k_2'}
G(s_1,s_2)_{k_2'j_2}\right]
\no \\ 
 & 
 \qquad \times F_\La[s_1,s_2]({\scriptstyle (i_1j_1), (i_2j_2) }) \ .
\end{align}
We repeat this argument until we construct all the possible
connected components containing the root cube.
This is a finite sum, for $\La$ fixed.

\begin{figure}
\centerline{\psfig{figure=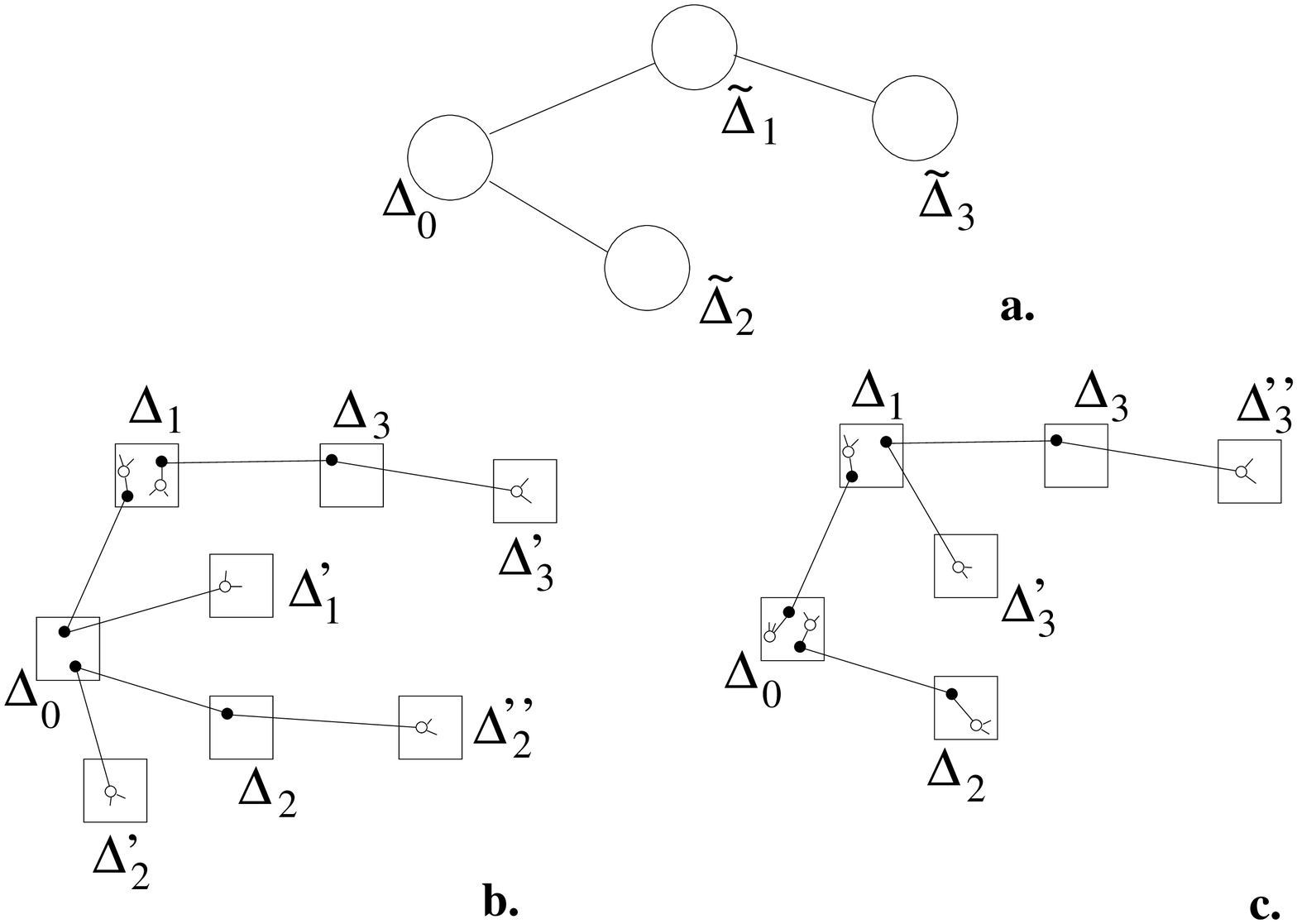,width=10cm}}
\caption{The two polymers and tree structures in {\bf b.} and  {\bf c.} 
both correspond to the same
 generalized polymer ${\tilde Y}$ and tree $T$ in {\bf a.}}
\label{tree}
\end{figure}

\paragraph{Definitions}
We give now some more precise statements. 
We define a {\em generalized cube}  ${\tilde \De}$ as a set of one, two or
three disjoint cubes in $\La$. A {\em generalized polymer} 
 ${\tilde Y}$ is then a  disjoint set of  generalized
cubes ${\tilde \De}$.  A tree $T$ on 
 ${\tilde Y}$ is a set of links $l_1$,.. $l_n$ connecting
the generalized cubes in  ${\tilde Y}$ and forming no loops (See Fig.\ref{tree}).
We call the set of all these cubes the polymer $Y$ contained in 
 ${\tilde Y}$.
Each link  $l_r$ corresponds to four 
connected vertices $i_r$, $j_r$, $k_r$, $k'_r$.
The corresponding propagators are $G_{i_r,k_r}$, $C_{k_r,k_r'}$
and  $G_{k'_r,j_r}$ defined in (\ref{Gdef}) and the corresponding
links are shown in  (see Fig.\ref{figlink}).
Note that the same  ${\tilde \De}$  and tree $T$ may correspond to
several different polymers $Y$ (see Fig.\ref{tree}).

We apply then the  following formula.

\begin{lemma}
$\left\langle G^+_{00}\right\rangle_{\vep =0} $ can be written as
\begin{align}
\left\langle G^+_{00}\right\rangle_{\vep =0} =& \sum_{{\tilde Y}\ni \De_0}\  
\sum_{T\, {\rm on} {\tilde Y} }\  \prod_{r=1}^{|{\tilde Y}|-1} 
 \int_0^1 ds_r
\sum_{\substack{(i_r,j_r)\\ (k_r,k'_r)}}   \, M_T(s) 
\label{cluster}\\
 &\left [  \prod_{r=1}^{|{\tilde Y}|-1}    G(s)_{i_r k_r}  
C_{k_rk_r'}
G(s)_{k_r'j_r}
 \right ]\, 
F_Y[s](\{i_r,j_r\})
\no 
\end{align}
where we sum over the generalized polymers ${\tilde Y}$ and over 
the ordered trees $T$ on  ${\tilde Y}$ with root $\De_0$.  
Note that the ordering on a tree  is the ordering on its links $l_r$,
$r=1,...|Y|-1$. The product over $r$ is then the product over the
links in the tree. The points  $(i_r,j_r)$ $(k_r,k'_r)$ fix the
number of cubes  inside each  ${\tilde \De}\in{\tilde Y}$ and
the links connecting them. 
Each tree link $l_r$ is associated to the parameter $0\leq s_r\leq 1$.
The product  $G_{i_rk_r}C_{k_rk'_r}G_{k'_rj_r}$
corresponds to  $\partial_{s_r} B(s_r)$ for each link $l_r\in T$
and the factor  $M_T(s)$ is the  product of $s$ factors
 extracted by the derivatives $\partial_r B(s_r)$.
Finally  $F_Y[s](\{i_r,j_r\})$ is the functional integral remaining 
after the propagators $\partial_{s_r} B(s_r)$  have been extracted.

More precisely, for each $r$ the points  $(i_r,j_r)$ $(k_r,k'_r)$
must satisfy  the constraint $k_r\in {\tilde\De}_{r'}$ for some
$r'<r$, $k'_r\in {\tilde\De}_{r}$, and 
$i_r, j_r$ may belong to any ${\tilde\De}_{r'}$
with  $r'\leq r$.  The propagators $C(s)$ $G(s)$ are defined as
\be
G(s) =: \frac{1}{1 +i m_i^2 C(s)},\quad \qquad C(s)_{ij}  =: s_{ij} C_{ij}
\label{Gdef}\end{equation}
where
\begin{equation}
\ba{ccccl}
s_{ii} &=& 1 & \qquad &   \\
s_{ij} &=& 1  &\qquad & {\rm if} \ \exists\  r \ {\rm s.t.}\ \  i,j\in {\tilde \De}_r \\
s_{ij} &=& \prod_{k=r}^{r'} s_k  &\qquad & {\rm if} \ \exists r'<r \ 
{\rm s.t.}\  i\in {\tilde \De}_r,\
j\in  {\tilde \De}_{r'}\\
s_{ij} &=& 0  & \qquad & {\rm if}\  i\in Y,\  j\not\in Y \ .\\
\ea
\end{equation}
The remaining functional integral is 
\begin{align}
& F_Y[s](\{i_r,j_r\})
= \, \int d\mu_{B(s)}(a,b,\rho^*,\rho )\   \label{FY} \\
&\ \prod_{r=1}^n
 \left [\frac{\de}{\de a_{i_r}} \frac{\de}{\de a_{j_r}} +
\frac{\de}{\de b_{i_r}} \frac{\de}{\de b_{j_r}} + 
\frac{\de}{\de \rho^*_{j_r}}  \frac{\de}{\de \rho_{i_r}}
\right ]  \, \left [e^{\sum_{j\in Y}(V_j-\rho^*_j\rho_j D_j)} \ {\cal O}_0
 \right ]
\no 
\end{align}
where $B(s)$ is
\be
B(s)^{-1} := C(s)^{-1} + i m_i^2 \ .
\end{equation}

Note that, as we constructed the tree, 
the order on the tree lines  ensures that, for each $1\leq r \leq n$, 
the tree line $l_r$ connects ${\tilde\De}_r$ to some
${\tilde\De}_{r'}$ with $r'<r$.
\end{lemma}

\paragraph{Proof}
The proof follows directly from the inductive procedure we explained above.
\qed \vskip 0.2cm

Now we can bound $|\langle G_{00}\rangle |$ inserting the
absolute value inside (\ref{cluster}).
Therefore we have to bound
\begin{align}
|\left\langle G^+_{00}\right\rangle_{\vep=0}| 
\leq & \sum_{{\tilde Y}}\  \sum_{T\, {\rm on} {\tilde Y}}\  
\prod_{i=1}^{|{\tilde Y}|-1} 
\left [ \int_0^1 ds_i
\sum_{\substack{(i_r,j_r)\\ (k_r,k'_r)}} \right ]  \, |M_T(s)| 
\label{cluster1}\\
 &\left [  \prod_{r=1}^{|{\tilde Y}|-1}    
| G(s)_{i_r k_r} | \,
|C_{k_rk_r'}|\,
|G(s)_{k_r'j_r}|
 \right ]\, 
|F_Y[s](\{i_r,j_r\})| \ .
\no 
\end{align}

\subsection{Bound on the  functional integral}
We bound the remaining functional integral $|F_Y[s](\{i_r,j_r\})|$ by a 
generalization of
the arguments for the finite volume case (Sec.5).

\paragraph{Definitions} 
We call ${\cal V}_d$ the set of vertices derived by the cluster
expansion appearing in (\ref{FY}):
\be
{\cal V}_d \ = \ \{ j\in Y\, : \, \exists\  1\leq r\leq |{\tilde Y}|-1\ {\rm s.t.}\ 
j=j_r\  {\rm or}\ j=i_r\}
\end{equation}
For each $j\in {\cal V}_d$ we call $d_j(a)$,  $d_j(b)$, $d_j(\rho)$, $d_j(\rho^*)$
the number of derivatives  $\de/\de a_j$, $\de/\de b_j$, 
$\de/\de \rho_j$ or $\de/\de \rho^*_j$ respectively.
As we see from (\ref{Vder}-\ref{detder2})
these derivatives apply to $V(a_j)$, $V(b_j)$ or $D_j\rho^*_j\rho_j$.
We also have a contribution from the observable ${\cal O}_0$, but only
for $j=0$.  

Note that, by the properties of anticommuting variables,
$d_j(\rho)$, $d_j(\rho^*)$ can be only 0 or 1. On the other hand there
is no limit to  $d_j(a)$,  $d_j(b)$. We call 
$d_j=d_j(a)+d_j(b)+d_j(\rho)+ d_j(\rho^*)$ the total number of derivatives
in $j$. This number is actually fixed by the choice of $\{(i_r,j_r)\}_r$.
For each $j\in {\cal V}_d$ we have to study 
\be
 \partial_{a_j}^{d_j(a)} \partial_{b_j}^{d_j(b)} 
\partial_{\rho^*_j}^{d_j(\rho^*)}
 \partial_{\rho_j}^{d_j(\rho)} \lp e^{V_j(a) + 
V_j(b)}(1 -\rho^*_j\rho_j D_j)\rp
\end{equation}
In the small field region we need to extract some structure. We can 
actually write the derivative as 
\begin{equation}
  \hspace{-0.2cm} 
\sum_{\substack{r_j(a),r_j(b)\\ r_j(\rho),r_j(\rho^*)} }\ 
a_j^{r_j(a)}\  b_j^{r_j(b)}\  \rho_j^{r_j(\rho)}\  {\rho_j^*}^{r_j(\rho^*)}\ 
C(a_j) C(b_j)
\end{equation}
where $r_j(a)$,  $r_j(b)$,  $r_j(\rho)$, $r_j(\rho^*)$ are
the number of fields remaining after the derivatives have been performed.
Note that  $r_j(\rho)$ and $r_j(\rho^*)$ can take only the values 0 and
1. On the other hand  it is easy to see that $r_j(a)\leq 3 d_j(b)$,
and the same holds for $b$. Moreover the parameter $n_j=d_j+r_j\geq 3$,
except for $j=0$ (and $j=x$ if we are considering $R(x)$).
The factors $C(a)$ and $C(b)$ no longer depend on 
the fermionic fields.
By  analytic tools we can show that in the small field region $I^1$
\be
\left |  C(a)\right | \leq K^{d_j}\ d_j(a)! \qquad 
\left |  C(b)\right | \leq K^{d_j}\  d_j(b)!
\end{equation}
In the large field region (that means in $I^q$ for $q>1$), 
we do not need to extract the whole
structure, as the fields are large and the small factors
come from the probability. We only need to extract explicitly the
fermionic fields therefore we write the derivatives as
\begin{equation}
  \hspace{-0.2cm} 
\sum_{ r_j(\rho),r_j(\rho^*) }\ 
 \rho_j^{r_j(\rho)}\  {\rho_j^*}^{r_j(\rho^*)}\ 
C(a_j) C(b_j)
\end{equation}
Again by analytic tools we can show that in the 
\be
\left| C(a) C(b)\right| \leq   K^{d_j}\ d_j(a)!
 d_j(b)!\ \left | e^{V_j(a)+V_j(b)}\right |
\end{equation}
In this region we have no factors $r_j$. Nevertheless we 
define $r_j=0$ if $d_j\geq 3$ and $r_j=3-d_j$ otherwise.
In this way we ensure $n_j\geq 3$ for all $j\in {\cal V}_d$.

With these definitions and results we prove the following Theorem.
\begin{theor}
\be
|F_Y[s](\{i_r,j_r\})| \ \leq \ K_1^{|Y|}\ 
\sum_{\{r_j\}_{j\in {\cal V}_d}}
\prod_{\De\in Y} 
\left[ K_2^{n_\De}\ r_\De!\  
\prod_{j\in {\cal V}_d\cap \De} d_j! 
\lp\frac{1}{W}\rp^{r_j}
\right]
\end{equation}
where $K_1$, $K_2$ are constants and  
$r_\De= \sum_{j\in {\cal V}_d\cap \De}r_j$,
$n_\De=\sum_{j\in {\cal V}_d\cap \De}n_j$. 
\end{theor}

\paragraph{Proof}
The proof is a  generalization of the arguments for Theorem~2 and 3 in 
Sec.~5.
Note that  Lemma~3 and 4  hold also after substituting $B$ and $C$ with $B(s)$
and $C(s)$.  Then we can introduce the partition (\ref{part}) in each cube 
separately
\be
1=\prod_\De \left [ \sum_{k_\De} \chi[I^{k_\De}_\De]\right]
\end{equation}
We perform the bounds in each cube as in Sec.~5. Note that, as the derivations
bring several bosonic and fermionic fields out of the exponential
we have to use some of the ideas of Theorem~3.

First we have to perform the integral over the fermionic variables 
extracting the correct factors $W^{-1}$.
Now we have many different cubes,
each in a different interval, therefore we cannot apply 
(\ref{fbound}) as in Lemma~9, as $(B^{-1}+D)^{-1}$ is not be well defined. 
We apply then a generalization of ((\ref{sbound}) which 
allows to extract also the small factors. 

We have to compute the integral
\begin{align}
& \int d\mu_{B(s)}(\rho^*,\rho )\  
 \left [ e^{-\sum_{j\in Y\backslash {\cal V}_d}\rho^*_j\rho_j (D_j+D'_0)}  
\right ] 
\label{fint} \\
&\quad \times\  \left [ \prod_{\substack {j\in {\cal V}_d\, :\, r_j(\rho)=1\\ 
r_j(\rho^*)=0 }} 
\hspace{-0.3cm} 
\rho_j \right ]  
\left [\prod_{\substack {j\in {\cal V}_d\, :\, r_j(\rho^*)=1\\ r_j(\rho)=0 }}
\hspace{-0.3cm}   
\rho^*_j   \right ] 
\left [  \prod_{\substack { j\in {\cal V}_d\, :\, \\ 
r_j(\rho)=r_j(\rho^*)=1 }}
\hspace{-0.3cm}  \rho_j \rho^*_j  \right ] \no
\end{align}
We partition now the set of $j\in Y$ as 
\bqa
{\cal U} &=& \{ j\in Y\ |\ j\not\in {\cal V}_d   \} \no\\
{ \cal D} &=& \{  j\in Y\ |\  j\in {\cal V}_d, \ r_j(\rho^*)=1\ {\rm or}\  
 r_j(\rho)=1\}
\eqa
Note that the points $j\in {\cal V}_d$ with $r_j(\rho^*)= r_j(\rho)=1$
do not give any contribution to the integral.
For each cube we introduce 
$r_\De(\rho)=\sum_{i\in {\cal V}_d\cap\De} r_j(\rho)$
and  $r_\De(\rho^*)=\sum_{i\in {\cal V}_d\cap\De} r_j(\rho^*)$.
Note that the fields $\rho_j$ are columns in the resulting matrix
and  the fields $\rho_j^*$ are rows
With these definitions we have the following Lemma.

\begin{lemma}
By multi-linearity of the determinant,  the fermionic integral (\ref{fint}) 
above can be written as
\be
\prod_\De \left [ \
\frac{r_\De(\rho)^{r_\De(\rho)} 
r_\De(\rho^*)^{r_\De(\rho^*)}}{W^{r_\De(\rho)+r_\De(\rho^*) }}\right ]
\si \ \det \, M
\end{equation}
where $\si$ is a sign depending on how we order the 
rows and column in  $M$, and $M$  is the block matrix
\begin{equation}
M = \lp \ba{cc}
M_{uu} & M_{ud} \\
M_{du} & M_{dd} \\
\ea
\rp
\label{Mr1}\end{equation} 
$M_{uu}$ is a matrix  corresponds to the elements
$(M)_{ij}$ with both $i,j\in {\cal U}$,
$M_{dd}$ to the  elements
$(M)_{ij}$ with  both $i,j\in {\cal D}$ 
and $M_{ud}$ and $M_{du}$ are the mixed terms. 
Note that, if $j\in {\cal V}_d$ and $r_j(\rho)=1$ but $r_j(\rho^*)=0$
$j$  appears  only
in  a column of $M_{dd}$ and the element $M_{jj}$ is not
present in the matrix. Therefore we order
the lines and columns of $M_{dd}$:   $(M_{dd})_{ij}= (M)_{l_i c_j}$ 
where  $l_i$ and $c_j\in  V_d$.
With these definitions the matrix elements are
\begin{equation}
\ba{ccccccc}
\lp M_{uu}\rp_{ij} &=& \left[ 1+ (D+D'_0) B(s)\right]_{ij}   &\quad&
\lp M_{ud}\rp_{ij} &=&  D_i B(s)_{i c_j} 
\frac{W}{ r_{\De(c_j)}(\rho)}   \\
\lp M_{dd}\rp_{ij} &=& \frac{W}{ r_{\De(l_i)}(\rho_j^*) } B(s)_{l_i c_j}  
\frac{W}{ r_{\De(c_j)}(\rho)} &\quad&
\lp M_{du}\rp_{ij} &=& \frac{W}{ r_{\De(l_i)}(\rho_j^*) }  B(s)_{l_ij} \\
\ea
\label{Mr2}
\end{equation} 
where $\De(i)$ is the cube containing the vertex $i$.
\end{lemma}
\paragraph{Proof}
The proof follows from the properties of the anticommuting
variables and determinants.
\qed\vskip 0.2cm

Now we can insert absolute values inside the bosonic integral.
If we bound $\det M$  as in (\ref{detb})  
we obtain the same bound as in Lemma~4, with an additional
error term. The precise statement is given in Lemma~(\ref{dbound})
below.

\begin{lemma}
The determinant of the matrix $M  $ 
defined as in (\ref{Mr1}-\ref{Mr2})  satisfies the bound
\begin{equation}
 |\det M| \ \lesssim\   e^{{\rm Re\, Tr} (M_{uu}-1) }
\  e^{O(|Y|+ |V_d| ) } 
\label{detbound}\end{equation}

\label{dbound}
\end{lemma}

\paragraph{Proof}
Using (\ref{detb})
we have 
\begin{equation}
 |\det M| \  \leq \   e^{{\rm Re\, Tr} (M-1) }
\  e^{K {\rm Tr} \lp M-1\rp^* \lp M-1\rp}
\end{equation}
Now applying the definitions 
(\ref{Mr1}) we can write 
\begin{align}
& {\rm Re\, Tr}\, (M-1)  ={\rm Re\, Tr}\, (M_{uu}-1) 
+ {\rm Re\, Tr}\, (M_{dd}-1) \no\\
& {\rm Tr} \,\lp M-1\rp^* \lp M-1\rp = 
  {\rm Tr} \, (M_{uu}-1)^*(M_{uu}-1) +
 {\rm Tr}\,  (M_{dd}-1)^*(M_{dd}-1) +\no\\
& \hspace{2cm}
{\rm Tr}  \,\left [(M_{du}-1)^*(M_{ud}-1) +
(M_{ud}-1)^*(M_{du}-1) \right ]
\end{align}

By inserting the definitions (\ref{Mr2}) and using the decay of
$B_{ij}$, it is easy to see that all terms are bounded by a constant
per cube except for
\be
 {\rm Tr}  (M_{dd}-1)^*(M_{dd}-1) \lesssim 
\sum_{\De} r_{\De}(\rho)+ r_{\De}(\rho^*) 
\end{equation}
This  completes the  proof. \qed
\vskip 0.2cm

Now we perform the estimates as in Sec.~5.
As in Sec.5.1.5, in order to decouple the estimates on different cubes  
we write all the bounds
in terms of quadratic forms like $v^TC_f(s) v$    
where $v$ is some vector and $C_f(s)= C(s)^{-1}- f m_r^2> 0$
as in (\ref{Cf})
but  $f$ is now a diagonal matrix which is constant on
each cube. Then  we can apply
\be 
C(s)\ <\ \frac{1}{-W^2 \De_N+m_r^2} 
\end{equation}
where $\De_N$ is the discrete Laplacian with Neumann boundary
conditions on the cubes. This operator decouples automatically
different cubes. Note that to estimate the contribution from
the small field region we have to apply (\ref{sfb1}) Lemma~10.
To extract the factors $W^{-r_j}$ in the large field cubes
we use the exponential factors $e^{-W}$. 

\subsection{Sum over the clusters}

We perform now the sum over the clusters $Y$. We split the sum 
in several pieces. First
 fixing the cubes $\De$ we sum over
the points $i_r$, $j_r$, $k_r$, $k'_r$ in the cubes.
Note that after this operation
is done, there is still a small factor associated to each cube. 
The factorials arising from the combinatorics
are beaten by a piece from the decay of $GCG$. 
The remaining piece of the decay is used to
sum over the cube positions, following the tree structure. 
Finally we sum over the tree choice $T$ using 
the fact that we have a small factor per cube.

To perform all these bounds we need now to study the spatial
decay of the propagators $C$, $B$ and $G$. We know already the 
spatial decay of $C(s)$ (see (\ref{Cdecay}). The decay of
$G$ is given by the
following Lemma.

\begin{lemma}
The propagator $G=(1+im_i^2 C(s))^{-1}$ decays as 
\be
|G_{ij}| \leq \de_{ij} + \tfrac{1}{W^2}\tfrac{ |m_i^2|}{1+|i-j|}
e^{-\frac{m_r}{W} |i-j|} +  \tfrac{2}{W^3}\tfrac{|m_i^4|}{m_r}
e^{-f\frac{m_r}{W} |i-j| } 
\end{equation}
where $f= \inf[ 1/2, g]$ and $ g$
is some constant independent from $W$. 
\end{lemma}

\paragraph{Proof} 
By a Combes-Thomas argument we prove that
\be
\left \|R^{-1} \frac{1}{1+im_i^2 C(s)} R \right  \| \leq 2
\end{equation}
for $R$ a multiplication operator defined as
$R|x> = \exp[\vec{\mu}\vec{x}]|x>$ and $\vec{\mu}$
any vector with $|\vec{\mu}|<g m_r/W$.
Now  $G$ can be written as
\be
\lp \frac{1}{1+im_i^2 C(s)}\rp_{ij}=
\de_{ij} -i m_i^2 C(s)_{ij} - m_i^4 
\lp  C(s) \frac{1}{1+im_i^2 C(s)}  C(s)\rp_{ij}
\end{equation}
We need to study the decay of the last term. 
Actually we see that
\begin{align}
& \left |  C(s) \frac{1}{1+im_i^2 C(s)}  C(s)\right |_{ij}
e^{\mu |\vec{x}-\vec{y}|}\  =\  
\left | R^{-1}  C(s) \frac{1}{1+im_i^2 C(s)}  C(s) R\right |_{ij}\no\\
&= \ \left | \lp R^{-1}C(s)R\rp \,
\lp R^{-1} \frac{1}{1+im_i^2 C(s)}R\rp\, \lp R^{-1}C(s)R\rp\right |_{ij}\no\\
&=\   \left | (V,A W)\right | \leq \|V \|\, \|W \|\, \|A\|
\ \leq \ \frac{2}{W^3}
\end{align}
where we defined $V_k= (R^{-1}C(s)R)_{ik}$,
 $W_j= (R^{-1}C(s)R)_{kj}$ and $A = R^{-1}(1+im_i^2 C(s))R$.
We chose  $\vec{\mu}$ such that 
$\mu |\vec{x}-\vec{y}| = \vec{\mu} (\vec{x}-\vec{y})$, and 
$\mu=|\vec{\mu}|\leq g m_r/W$,
and  $\mu\leq m_r/2W$. The last condition must ensure
that the exponential decay of $C(s)$ controls the exponentials
from $R^{-1}$ and $R$.
This completes the proof. \qed
\vskip 0.2cm

Note that this Lemma gives also the decay of $B(s)$
\be
|B(s)_{ij}| \leq \tfrac{1}{W^2}\tfrac{1}{1+|i-j|}
e^{-\frac{m_r}{W} |i-j|} +  \tfrac{2}{W^3}\tfrac{|m_i^2|}{m_r}
e^{-f\frac{m_r}{W} |i-j| }
\end{equation}
as $B(s)=(G-1)i/m_i^2$.

\subsubsection{Extracting small factors}                                        
Before performing the estimates we extract some  factors from 
the propagators $GCG$ for each tree line, to offset the factorials 
eventually arising in the estimates and to ensure we have 
a small factor  per vertex.

\paragraph{Factorials.} Constant powers of factorials such as $d_\De^p$,
for $p$ fixed, can be beaten using a piece of the decay of $CGC$. 
Note that each tree line $l_r$ connects {\em different} cubes, therefore we have
$d_\De$ disjoint cubes hooked to the cube $\De$ by the tree $T$.
When  $d_\De$ is large, since we are in a finite dimensional space, many of these cubes
must be very far from $\De$. It is easy to see that, half of the 
 $d_\De$ cubes must be at a distance from $\De$ of order $W d_\De^{\frac{1}{3}}$.
Therefore we gain a factor $\exp[- c\,d_\De^{\frac{4}{3}} ]$ which can beat any
constant power of factorials. 
\be
\left [  \prod_{r=1}^{|{\tilde Y}|-1}   
e^{-\vep\frac{|i_r-k_r|}{W}} e^{-\vep\frac{|k_r-k'_r|}{W}} 
e^{-\vep\frac{|k'_r-j_r|}{W}} 
\right ] \leq 
 \left [\prod_{\De}  
\frac{K^{n_\De}}{n_\De!^p}\right ] 
\label{fact}\end{equation}
Note that the constant $K$ depends on $p$.

\paragraph{$W$ factors} We need  a factor $W^{-1}$ for each field
hooked to a derived vertex $j\in {\cal V}_d$.  We extract then a factor $W^{-1}$ 
from each $G$
propagator:
 \begin{align}
&\hspace{-0.5cm}\left [  \prod_{r=1}^{|{\tilde Y}|-1}   
| G(s)_{i_r k_r} | \,
|G(s)_{k_r'j_r}| 
 \right ]\, \left [  \prod_\De \lp\frac{1}{W}\rp^{r_\De}\right ]\\
& = \left [  \prod_{r=1}^{|{\tilde Y}|-1}   
| WG(s)_{i_r k_r} | \,
|WG(s)_{k_r'j_r}| 
 \right ]\,
\left [\prod_{j\in {\cal V}_d} 
\lp\frac{K}{W}\rp^{n_j}\right ] \no
\end{align}
Note that $n_j\geq 3$ for all $j\in {\cal V}_d$ except for $j=0$. As
the each cube has volume $W^3$, this ensure
that we can choose the position of each vertex without 
paying any factor $W$.

Finally, after extracting  a fraction
$\vep$ of the exponential decay and the factors $W^{-1}$ 
we separate in the remaining factors  the polynomial and exponential decay
\begin{equation}
 \left [  \prod_{r=1}^{|{\tilde Y}|-1}   
 {\tilde G}_{i_r k_r}  \, {\tilde C}_{k_r, k'_r}\, 
 {\tilde G}_{k'_r j_r}\right ]
\left [  \prod_{r=1}^n   e^{-\frac{f'}{W}d(\De_{(i_r)},\De_{(k_r)})}
e^{-\frac{f'}{W} d(\De_{(k_r)},\De_{(k'_r)})}
 e^{-\frac{f'}{W} d(\De_{(k'_r)},\De_{(j_r)})} \right ]
\label{gendecay}
\end{equation}
where 
\be
 {\tilde G}_{i j} =: 
  \left [\de_{i_rk_r} W^2 + 
\tfrac{1}{1+|i-j|}+ \tfrac{1}{W} \right ] \qquad
 {\tilde C}_{i j} =:  \tfrac{W^{-4}}{1+|k_r-k'_r|}\
\end{equation}
where $f'=fm_r-\vep$ is the remaining mass and $d(\De,\De')$ is the
distance between the center of the cube $\De$ and $\De'$.

\subsubsection{Sum over the vertex positions}
Now using the decay $1/|i-j|$ in ${\tilde G}$ and  ${\tilde C}$
we sum over the positions of
all vertices  inside their cube (the cube is fixed).
Each line of the cluster expansion corresponds to four vertices
$i_r,j_r,k_r,k'_r$, where $k_r$ and $k'_r$ correspond to
two point vertices and must belong to different cubes while 
$i_r$, $j_r$ may belong to the same cube. For each $j=i_r$ or
$j=j_r$ we distinguish two cases
\begin{itemize}
\item{} $j$ contracts to $j'$ and $j'$ has never been extracted before
in the cluster expansion. 
Then we say $j'$ is new with respect to $j$.
\item{} $j$ contracts to $j'$ and $j'$ has already been extracted before
by the cluster expansion. Then we say $j'$ is old
with respect to $j$.
\end{itemize}

We consider first the case $k_r\neq i_r$ and
  $k'_r\neq j_r$ so that the factors $\de_{i_rk_r}$ and
 $\de_{j_rk'_r}$ disappear. 
Note that we sum over the position of  $i_r$ ($j_r$) only when it is new.
We consider the different cases.

\paragraph{$i_r\neq j_r$ and both  $i_r$ and $j_r$ new} Then we sum 
over $i_r$ and  $j_r$.
\begin{equation}
\sum_{k_r\in \De_{(k_r)}} \sum_{i_r\in \De_{(i_r)}} 
 {\tilde G}_{i_r k_r}
 \sum_{k'_r\in  \De_{(k'_r)}}   {\tilde C}_{k_r, k'_r}\, 
\sum_{j_r\in \De_{(j_r)}} {\tilde G}_{k'_r j_r}
\leq W^{5}\  =\  \lp W^{\frac{5}{2}}\rp \  \lp W^{\frac{5}{2}}\rp 
\end{equation}
Therefore we pay a factor $W^{5/2}$ for $i_r$ and the same factor 
for $j_r$.

\paragraph{$i_r\neq j_r$ and   $i_r$ new,  $j_r$ old} 
Then we sum only over $i_r$.
\begin{equation} 
  \sum_{k'_r\in \De_{(k'_r)}}  
 {\tilde G}_{k'_r j_r}
\,  \sum_{k_r\in  \De_{(k_r)}}  {\tilde C}_{k_r, k'_r}\, 
 \sum_{i_r\in \De_{(i_r)}} 
 {\tilde G}_{i_r k_r}
 \leq W^2
\end{equation}
Therefore we pay a factor $W^2$ for $i_r$ and no factor  
for $j_r$.

\paragraph{$i_r\neq j_r$ and   
$i_r$ and  $j_r$ old} Then $i_r$ and $j_r$ are both 
fixed.
\begin{equation}
  \sum_{k_r\in \De_{(k_r)}}
{\tilde G}_{i_r k_r} W^{-4}
 \sum_{k'_r\in  \De_{(k'_r)}}  
{\tilde G}_{k'_r j_r}
\leq\  O(1)
\end{equation}
where we bounded $ {\tilde C}_{k_r, k'_r}\leq W^{-4}$.

\paragraph{$i_r= j_r$ and   $i_r$ new} Then we sum over $i_r$.
\begin{equation}
 \sum_{k_r\in \De_{(k_r)}} 
 \sum_{k'_r\in  \De_{(k'_r)}} 
 {\tilde C}_{k_r, k'_r}
\left \{\sum_{\substack{ i_r\in \De_{(i_r)}\\ |i_r-k_r|\geq  |i_r-k'_r|}}
\left [{\tilde G}_{i_r k'_r}\right ]^2
+\hspace{-0.2cm} \sum_{\substack{i_r\in \De_{(i_r)}\\ |i_r-k_r|< |i_r-k'_r|}}
\left [{\tilde G}_{i_r k_r}\right ]^2\right \}
\leq W^2 
\end{equation}
Therefore we pay a factor $W^2$ for $i_r$.

\paragraph{$i_r= j_r$ and   $i_r$  old} Then $i_r$ is fixed.
\begin{equation}
 \sum_{k_r\in \De_{(k_r)}} 
{\tilde G}_{i_r k_r}  W^{-4}
 \sum_{k'_r\in  \De_{(k'_r)}} 
{\tilde G}_{k'_r j_r}
\ \leq\  O(1)
\end{equation}
where we bounded  $ {\tilde C}_{k_r, k'_r}\leq W^{-4}$.
When $i_r=k_r$ or  $j_r=k'_r$ it is easy to see that the same 
estimations hold.

Note that for each $j=i_r$ ($j=j_r$), with $j\neq 0$, we pay a factor
$W^{5/2}$  when it is new and some  constant $K$ in any other case.
Therefore, applying  $n_j\geq 3$, we have
\begin{equation}
W^{\frac{5}{2}} K^{n_j-1} \frac{1}{W^{n_j}} = 
\lp\frac{K}{W^{1-\frac{5}{2n_j}}}\rp^{n_j} 
\leq \lp\frac{K}{W^{\frac{1}{6}}}\rp^{n_j}
\label{count}\end{equation}
Therefore we have a factor $W^{-\frac{1}{6}}$ for each tree line hooked to $j$.
This means we have a factor  $W^{-\frac{1}{6}}$ for each generalized cube
${\tilde \De}$ is ${\tilde Y}$.
The case  $j= 0$ is special as $n_j\geq 0$. Nevertheless
the position of 0 is fixed so that we do not pay the factor
$W^{5/2}$. Therefore for $j=0$ we have 
\begin{equation}
 K^{n_0} \frac{1}{W^{n_0}} 
\label{count1}\end{equation}

\subsubsection{Combinatorial bounds}

The combinatoric inside each cube costs a factor 
\be
\sum_{r_\De\leq 3 d_\De} K^{d_\De} (d_\De!) (r_\De!) \leq   K^{d_\De} (d_\De!)^4 =
 K^{|{\tilde Y}|} (d_\De!)^4 
\end{equation}
The factorials are beaten by a piece of the exponential
decay of the tree lines (\ref{fact}), while the constant 
factor will be bounded later when we will sum over the
tree structure.

\subsubsection{Sum over the cube positions and the tree structure}

We sum over the cube positions using the exponential
decay and following the tree from the leaves towards the root.
The result is $K^{|{\tilde Y}|}$.
The remaining sum  is now
\begin{equation}
 |\left\langle G^+_{00}\right\rangle_{\vep=0}|  \lesssim  
\sum_{|{\tilde Y}|}\  \sum_{\substack{T \\ {\rm unordered}}}\ 
\sum_{{\rm orders }} \prod_{i=1}^{|{\tilde Y}|-1} 
 \int_0^1 ds_i \, |M_T(s)|
\prod_{{\tilde \De\in {\tilde Y}}} \left [  \frac{K}{W^{\frac{1}{6}}} 
\right ] 
\end{equation}
where we have split the sum over  ordered rooted trees as 
the sum over  unordered rooted trees, and the sum over orders.
This last sum is performed by the integral over the
interpolating factors (see \cite{Bry} or \cite{R}, Lemma III.1.1)
We give here a sketch of the proof.
\begin{lemma}
The sum over all the orders on the tree $T$ 
is bounded using the interpolating factors $s_i$ as follows
\begin{equation}
  \sum_{\rm orders} \prod_{i=1}^{|{\tilde Y}|-1}  
 \int_0^1 ds_i \, |M_T(s)| = 1
\label{Mt}\end{equation}
\end{lemma}

\paragraph{Proof}
We introduce the variables $\vep_{ij}$ for all ${\tilde\De}_{i}$
 ${\tilde\De}_{j}\in {\tilde Y}$. Then we introduce the function
$F(\vep)=: \prod_{(ij)\in T} (1+\vep_{ij})$ where $T$ is unordered.
Now we perform the tree expansion as we did in Lemma~12 Sec.~6.1.
We define $\vep_{ij}(s_1)=s_1\vep_{ij}$ if $i$ or $j=0$ and
 $\vep_{ij}(s_1)=\vep_{ij}$ otherwise. We apply the first order Taylor expansion
and we go on until we extract all the $\vep_{ij}$.
The term proportional to all  $\vep_{ij}$ is then
\be
 \prod_{(ij)\in T}\vep_{ij}  \sum_{\rm orders} \prod_{i=1}^{|{\tilde Y}|-1}  
 \int_0^1 ds_i \, |M_T(s)| 
\end{equation}
If we expand $F(\vep)$ in powers of $\vep$ we see that the
term $ \prod_{(ij)\in T}\vep_{ij} $ has coefficient 1.
Therefore by comparing powers of $\vep$ we obtain (\ref{Mt}).
\qed
\vskip 0.2cm

Finally the sum over the structure  can be written as 
\begin{equation}
\sum_{|{\tilde Y}|}\  \sum_{\substack{T \\ {\rm unordered}}}\ 
 g^{|{\tilde Y}|}
 = 
\sum_{d_{{\tilde\De}_0}\geq 0} \prod_{i=1}^{d_{{\tilde\De}_0}}
g \left [
\sum_{d_{{\tilde \De}_i}\geq 0}
\prod_{i'=1}^{d_{{\tilde\De}_{i}}}
g \sum  \hdots 
\right ]
\end{equation}
where we defined $g=K/W^{\frac{1}{6}}<1$ 
and for each generalized cube we sum over the coordination
number. We start summing from the leaves going towards the root.
The leaves give
\be
g \sum_{d\geq 0} g^d = g \frac{1}{1-g} \leq g^{\frac{1}{2}},
\qquad  {\rm if }\ \  g^{\frac{1}{2}} \frac{1}{1-g}<1
\end{equation}
The following step gives
\be
g \sum_{d\geq 0} g^\frac{d}{2} = g \frac{1}{1-g^{\frac{1}{2}}} 
\leq g^{\frac{1}{2}},
\qquad {\rm if }\ \   g^{\frac{1}{2}} \frac{1}{1-g^{\frac{1}{2}}}<1
\end{equation}
Then we repeat inductively.
Finally 
\begin{equation}
\sum_{|{\tilde Y}|}\  \sum_{\substack{T \\ {\rm unordered}}}\ 
 g^{|{\tilde Y}|} \leq K + 1
\end{equation}
where the constant comes from the bound for $n=0$. 
This completes the proof of the first part of Theorem~4,
namely the boundness of the density of states (\ref{rhob}).

\subsection{Smoothness and exponential decay}

To bound the derivatives of the density of states (\ref{derivn}), and in 
particular the decay of $R(x)$ (\ref{Rdec}), we perform the cluster expansion
as in Sec.~6.1. 

For $R(x)$, note that contributions where $Y$ does not 
contain both 0 and $x$ are cancelled. When 0 and  $x$ 
both belong to $Y$ we can extract the exponential decay 
directly from the tree lines $GCG$.
For derivatives $\partial_E^n {\bar \rho}_\La(E)$ the idea is the same. 
Only contributions from $Y$ containing all the observables are not cancelled. 
The fine structure (the factor $W^{-3}$ in $R(x)$) are then extracted
by a few steps of perturbative expansion as in Sec.~5.2.2.
In the same way we can prove the semicircle law behavior (\ref{semic}).
This completes the proof of Theorem~4.

\vskip 1cm

\resetsect 

\renewcommand{\thesection}{\Alph{section}}

\noindent{\Large {\bf Appendix A }}
\medskip

\noindent{\large {\bf Supersymmetric formalism}}
\vskip 0.5cm

\resetequ

We summarize the conventions and notations we adopted in this work
(they are based on  the review by Mirlin \cite{M}). 

\paragraph{Grassmann variables} A set of Grassmann variables
and their complex conjugates 
 $\chi_1$, $\chi^*_1$, ...$\chi_N$ $\chi^*_N$ has the following
properties:
\be
\chi_i \chi_j = - \chi_j\chi_i,  \qquad   \qquad 
\chi^*_i \chi_j = - \chi^*_j\chi_i,
 \qquad   \qquad \chi^*_i \chi^*_j = - \chi^*_j\chi^*_i 
\label{chi1}\end{equation}
\be
 (\chi_i^*)^* = - \chi_i,   \qquad   \qquad 
(\chi_i\chi_j)^*= \chi_i^* \chi_j^* 
\label{chi2}\end{equation}
\be
 \int d\chi_i\  1 =  \int d\chi^*_i\  1 = 0,  \qquad 
  \qquad   \int d\chi_i \ \chi_i =  \int d\chi^*_i\  \chi^*_i = 
\frac{1}{\sqrt{2\pi}} 
\label{chi3}\end{equation}
With these definitions we introduce a vector and its adjoint as
 usual
\be
\chi =\lp \ba{c}
\chi_1 \\
\vdots \\  
\chi_N \\
\ea\rp  \qquad   \qquad 
\chi^+ = \lp \chi_1^*, \cdots , \chi_N^* \rp
\end{equation}
Now $\chi^+\chi$ is a real
commuting variable and 
\be
\int \prod_i   d\chi^*_i d\chi_i\ e^{- \chi^+ M \chi }
= \det \frac{M}{2\pi}
\end{equation}
for any matrix $M$. 

\paragraph{Supervectors and supermatrices}

A supervector is defined as
\be
\Phi = \lp \ba{c}
S_1 \\
\vdots \\  
S_N \\
\chi_1 \\
\vdots \\  
\chi_N \\
\ea\rp  \qquad   \qquad 
\Phi^+ = \lp  S_1^*, \cdots , S_N^*, \chi_1^*, \cdots , \chi_N^* \rp
\end{equation}
where $S_i$ are the commuting and $\chi_i$ are the anticommuting
components.
Similarly a supermatrix is a matrix with both commuting and 
anticommuting entries
\be
M =  \lp \ba{cc}
a & \si \\
\rho & b \\
\ea\rp
\label{sm}\end{equation}
where $a$ and $b$ are ordinary matrices while $\si$
and $\rho$ have anticommuting elements.
We identify the element of a supermatrix by four indices
$M_{ij}^{\al\bt}$ where $\al,\bt$ specify in which
sector we are: (0,0) corresponds to $a$
(boson-boson);  (1,1) corresponds to $b$ 
(fermion-fermion);
(0,1) corresponds to $\si$ 
(boson-fermion); (1,0) corresponds to $\rho$ 
(fermion-boson). $(i.j)$ identify the matrix element
inside each sector. For example
$M^{00}_{i j}= a_{ij}$. 

The notions equivalent to trace and determinant are
supertrace and superdeterminant
\be
{\rm Str} M = {\rm Tr} a -   {\rm Tr} b, 
\qquad \qquad {\rm Sdet} M = \det (a - \si b^{-1} \rho)\  
\det b^{-1}
\end{equation}
With these definitions we have 
\be
{\rm Str} \ln M =  \ln{\rm Sdet} M
\end{equation}
\be
\int d\Phi^*  d\Phi \ e^{-\Phi^+ M \Phi} =  {\rm Sdet} M^{-1}
\end{equation}
\be
\int d\Phi^*  d\Phi \ \Phi_{\al, k} \Phi_{\bt, l}^*\  
 e^{-\Phi^+ M \Phi} = ( M^{-1})^{\al \bt}_{kl} \  {\rm Sdet} M^{-1}
\end{equation}
Note that some properties are different from that of
the usual matrices, in particular:
\be
 {\rm Sdet}\  z M = {\rm Sdet} M 
\end{equation}
for any complex number $z$. 

Finally from these formulas one can derive the inverse of
the supermatrix $M$ (\ref{sm}):
\be
M^{-1} = \lp \ba{cc}
\lp a - \si b^{-1} \rho \rp^{-1} & 
-\lp a - \si b^{-1} \rho \rp^{-1}\si b^{-1} \\
- b^{-1} \rho \lp a - \si b^{-1} \rho \rp^{-1} & 
b^{-1} \left [
1 + \rho \lp a - \si b^{-1} \rho \rp^{-1} \si b^{-1}
\right ] \\
\ea\rp
\end{equation}


\vskip.2cm





\begin{thebibliography}{99}

\bibitem{M} A. Mirlin;
  Statistics of energy levels and eigenfunctions
in disordered and chaotic systems: Supersymmetry approach.
Phys. Rep. {\bf 326} (2000) 259 cond-mat/0006421 


\bibitem{AOS} A. Atland, C. R. Offer, B. D. Simons; in
{\em Supersymmetry and Trace Formulae. Chaos and Disorder.},
pg. 17,  edited  by
I.V. Lerner, J.P.Keating and D.E. Khmelnitskii (1999)


\bibitem{BA} P. M. Bleher, A. R. Its editors; 
{\em Random Matrix Models and their Applications},
MSRI Publications, Vol. {\bf 40} (2001)


\bibitem{KS} N. M. Katz, P. Sarnak;
Zeroes of Zeta functions and symmetry,
Bull. of the Am. Math. Soc. Vol. {\bf 36} No. 1 (1999) 1  


\bibitem{J} N. K. Johansson;
Shape fluctuations and Random matrices,
Comm. Math. Phys.  {\bf 209} (2000) 437  


\bibitem{PF} L. Pastur, A. Figotin;
{\em Spectra of Random and Almost-Periodic
Operators} Springer-Verlag
 

\bibitem{AS} M. Aizenman, J. Schenker, R. M. Friedrich, D. Hundertmark;
Finite-Volume Fractional-Moment Criteria for Anderson Localization,
to appear in Comm.  Math. Phys. 


\bibitem{W}  F. J. Wegner; 
 The Mobility Edge Problem: Continuous Symmetry and a Conjecture,
 Z. Phys.   {\bf B 35} (1979) 207

\bibitem{SW} L. Sch\"afer, F. J. Wegner; 
Disordered System with n Orbitals per Site: Lagrange Formulation, 
Hyperbolic Symmetry, and Goldstone Modes, 
 Z. Phys.   {\bf B 38} (1980) 113


\bibitem{MF} Y. V. Fyodorov, A. Mirlin; 
Scaling properties of Localization in Random Band Matrices: 
A $\sigma$-Model Approach, 
 Phys. Rev. Lett.   {\bf 67} no. 18 (1991) 2405 


\bibitem{Ef} K. B. Efetov; 
Supersymmetry and theory of disordered metals,
Adv. Phys.  {\bf 32} no. 1 (1983) 53

\bibitem{Ef1} K. B. Efetov; {\em Supersymmetry in Disorder and Chaos},
Cambridge University Press  1997


\bibitem{Kl} A. Klein; 
 The supersymmetric replica trick and smoothness of the density of
states for random Schr\"odinger operators, 
in {\em Operator theory: operator algebras and applications}
Part 1, pg 315; Proceedings of symposia in pure mathematics  {\bf 51}, part 1 


\bibitem{BL} H. J. Brascamp, E. H. Lieb;
 On extension of the Brunn-Minkowski and 
Pr\'ekopa-Leindler theorems, including
inequalities for log concave functions and 
with an application to the diffusion equation,
J. of Func. An. {\bf 22} (1976) 366  

\bibitem{Sp} T. Spencer;
 Scaling, the free field and statistical mechanics,
Proc. of Symp. in Pure Math. {\bf 60} (1997) 373  



\bibitem{Bry} D. Brydges ; 
 {\em A short course on cluster expansions}, Les Houches
session XLIII, 1984, Elsevier Sience Publishers (1986)

\bibitem{R} V. Rivasseau; {\em From perturbative to contructive
renormalization}, Princeton Series in Physics
\vskip.1cm











\end{thebibliography}
\end{document}